%% file: metagenBERT-mcardis.tex
\documentclass[long]{doc}

\RequirePackage[default,scale=0.90]{opensans}
\usepackage[utf8]{inputenc}
\usepackage{url}
\usepackage{makecell}
\usepackage{multirow, booktabs}
\usepackage{subcaption}

\graphicspath{{fig/}}

\title{MetagenBERT: a Transformer-based Architecture using Foundational genomic Large Language Models for novel Metagenome Representation}

\author{
  Gaspar \uppercase{Roy}\inst{1}
  \and
  Eugeni \uppercase{Belda}\inst{1,2}
  \and
  Baptiste \uppercase{Hennecart}\inst{1}  
  \and
  Yann \uppercase{Chevaleyre}\inst{3}
  \and
  Edi \uppercase{Prifti}\inst{1,2}
  \and
  Jean-Daniel \uppercase{Zucker}\inst{1,2}}

\institute{
 IRD, Sorbonne University, UMMISCO, 32 avenue Henry Varagnat, Bondy Cedex, France
 \and
 Sorbonne University, INSERM, Nutriomics, 91 bvd de l'hopital 75013 Paris, France
 \and
 LAMSADE, Dauphine University, PSL Research University, Place du Maréchal de Lattre de Tassigny, Paris, France
}

\corresponding{gaspar.roy@ird.fr}

\papername{Roy \textit{et al.} (2025) MetagenBERT: a Transformer Architecture using Foundational DNA Read Embedding Models to enhance Disease Classification}

\abstract{ 
Metagenomic disease prediction commonly relies on species-abundance tables derived from large and incomplete reference catalogs, a strategy that constrains resolution, increases computational costs, and discards valuable information contained in raw sequencing reads. To overcome these limitations, we introduce MetagenBERT, a Transformer-based framework that produces end-to-end metagenome embeddings directly from raw DNA sequences, without taxonomic or functional annotations. Individual reads are embedded using foundational genomic language models (DNABERT-2 and the microbiome-specialized DNABERT-MS), then aggregated through a scalable global clustering strategy based on FAISS-accelerated K-Means. Each metagenome is represented as a cluster-abundance vector summarizing the distribution of its embedded reads.

We evaluate this approach on five benchmark gut microbiome datasets (Cirrhosis, T2D, Obesity, IBD, CRC) and on the large, phenotypically diverse MetaCardis cohort (n = 2{,}022). MetagenBERT achieves competitive or superior AUC performance relative to species-abundance baselines across most tasks. Concatenating species abundances with embedding-based cluster abundances further improves prediction, demonstrating complementarity between taxonomic and embedding-derived signals. Global clustering remains robust when applied to as little as 5--10\% of reads, highlighting substantial redundancy in metagenomes and enabling major computational gains.

We additionally introduce MetagenBERT-Glob-MCardis, a cross-cohort variant in which clusters are trained on MetaCardis and transferred to benchmark datasets. Despite reduced performance compared to dataset-specific clustering, the MetaCardis-trained clusters retain strong predictive signal---including for phenotypes absent from MetaCardis---indicating the feasibility of a foundation model for metagenome representation. Robustness analyses (PERMANOVA, PERMDISP, entropy metrics) show consistent separation of healthy and diseased states and stable cluster structures across subsamples.

Overall, MetagenBERT provides a scalable, annotation-free, and interpretable representation of metagenomes, bridging foundational genomic language models with population-level microbiome variation. These results point toward future phenotype-aware metagenomic LLMs capable of generalizing across heterogeneous cohorts and sequencing technologies.
}

\keywords{Metagenomics, Gut microbiome, Transformer models, DNA sequence embedding, Disease classification}

\usepackage{setspace}
\begin{document}

\selectlanguage{english}

\maketitle

\section{Introduction}
\label{sec:intro}

\input{LaTeX/sections/1.Introduction/1.Introduction}

\section{Materials and Methods}
\label{sec:materials_and_methods}

\input{LaTeX/sections/2.Materials_And_methods/2.Materials_and_Methods}

\section{Results}
\label{sec:results}

\input{LaTeX/sections/3.Results/3.Results}

\section{Discussion}
\label{sec:discussion}

\input{LaTeX/sections/4.Discussion/4.Discussion}

\section{Data and Code Availability}

DNA sequences are available from EBI : ERP005860 for cirrhosis, ERP003612 for obesity, ERP005534 for CRC, ERA000116 for IBD and NCBI (SRA045646 and SRA050230) for type 2 diabetes. The code can be found on Github following this link : \url{https://github.com/CorvusVaine/MetagenBERT}

\section*{Funding}

This work was supported by a grant from the French "Agence Nationale de la Recherche" (ANR) for the DeepIntegrOmics project number ANR ANR-21-CE45-0030.

\section*{Acknowledgments}

This work was granted access to the HPC resources of IDRIS under the allocations 2023-AD011014580, 2024-AD011014580R1 and 2024-AD011015723R1 made by GENCI. \\
This work was supported by a grant from the French "Agence Nationale de la Recherche" (ANR) for the DeepIntegrOmics project number ANR ANR-21-CE45-0030.

\section{Conflict of interest disclosure}

The authors declare having no financial conflicts of interest in relation to the content of the article.

\bibliography{biblio}

\end{document}

%% file: LaTeX/sections/1.Introduction/1.Introduction.tex
Traditional metagenome representations typically rely on abundance tables, which serve as baseline embeddings of microbial communities \cite{basics}. While effective for many tasks, such approaches face several limitations: they are sensitive to taxonomic resolution, incur high computational costs, and may oversimplify microbial diversity \cite{knowns}. Furthermore, they depend on large yet incomplete reference catalogs, preventing them from serving as  foundation models : as reference databases evolve, embeddings must be recalculated and models retrained, hindering scalability and long-term consistency. \cite{unknown-microbes} Moreover, by reducing sequences to species counts, they lose valuable information about functional potential and ecological roles within the microbiome. \cite{wu_guild-based_2021}

To this date, these representations are used by Machine Learning (ML) and Deep Learning (DL) models to classify gut metagenomic samples from patients with various diseases such as Cirrhosis, Colorectal Cancer or Type 2 Diabetes. \cite{pasolli_machine_2016}\cite{reiman_popphy-cnn_2020}\cite{oh_deepmicro_2020}

In the meantime, relying on the analogy between DNA and Natural Language, Large Language Models (LLM) architectures have been adapted to biological sequences, and in particular to genomic sequences, resulting in the new class of genomic Large language Models (gLLMs) \cite{gLLM}.
Through self-supervised pretraining, these models learn regulatory grammar and sequence structures into mathematical representations named embeddings with models such as DNABERT-2. \cite{zhou2024dnabert2}
They act as genomic foundation models and can be fine-tuned to produce representations adapted to various tasks such as gene annotations, variant predictions or taxonomic classification. \cite{zhou2O24dnabertS} \cite{dalla-torre_nucleotide_2025}

However, these models still face several limitations. Benchmark studies show mixed performance across core genomic tasks, highlighting limitations in sequence alignment accuracy and reasoning without external reference integration. \cite{geneturing}

More recent models focus on integrating longer sequences potentially hoarding more expressive power in order to produce more robust representations, with models such as Gene42 or EVO. \cite{gene42} \cite{evo}

Lastly, metagenomic-based LLMs have very recently emerged. Trained specifically for pathogen detection (METAGENE-1 \cite{metagene-1}) or metagenomic annotation (GENERanno \cite{generanno}), they open the way to specialized metagenomic-LLM.

To address the challenges coming with species-based representations, we introduce \textbf{MetagenBERT}, a deep learning architecture that directly processes raw sequencing reads to produce end-to-end metagenome embeddings. Inspired by advances in Natural Language Processing, our approach integrates the gLLM models' expressive power in metagenomic classification to extract rich biological information from sequence data and propose a new representation of metagenomic samples without relying on external annotations or reference databases.

%% file: LaTeX/sections/2.Materials_And_methods/2.Materials_and_Methods.tex
\subsection{Pipeline Overview}
\label{subsec:pipeline}
\input{LaTeX/sections/2.Materials_And_methods/2.1.Pipeline_overview}

\subsection{Datasets}
\label{subsec:datasets}
\input{LaTeX/sections/2.Materials_And_methods/2.2.Datasets}

\subsection{Read Embedding}
\label{subsec:read}
\input{LaTeX/sections/2.Materials_And_methods/2.3.Read_embedding}

\subsection{Metagenome Embedding}
\label{subsec:metagenome}
\input{LaTeX/sections/2.Materials_And_methods/2.4.Metagenome_Embedding}

%% file: LaTeX/sections/2.Materials_And_methods/2.1.Pipeline_overview.tex
Our method processes a metagenomic sample—typically a FASTA or FASTQ file containing millions of short, unordered DNA reads from multiple species—to produce a single embedding that represents the entire metagenome. Unlike approaches relying on taxonomic annotations, our goal is to generate these embeddings directly from raw reads in an end-to-end manner.

Conceptually, a metagenome can be viewed as an unordered collection of short “sentences,” analogous to a massive, unstructured text. However, unlike natural language data, metagenomic reads lack sequential order, exhibit vast diversity across species, and may overlap or share conserved regions, posing unique computational challenges.

Our pipeline comprises two main stages: (1) read embedding and (2) embedding aggregation. In the first stage, a large language model encodes each read into a vector representation. In the second, these read embeddings are aggregated into a single fixed-length vector summarizing the entire sample for downstream classification. Details of these components and the datasets used for evaluation are presented in later sections.

\begin{figure}[!htp]
   \centering
    \begin{subfigure}[b]{\textwidth}
        \centering
        \includegraphics[width=\textwidth]{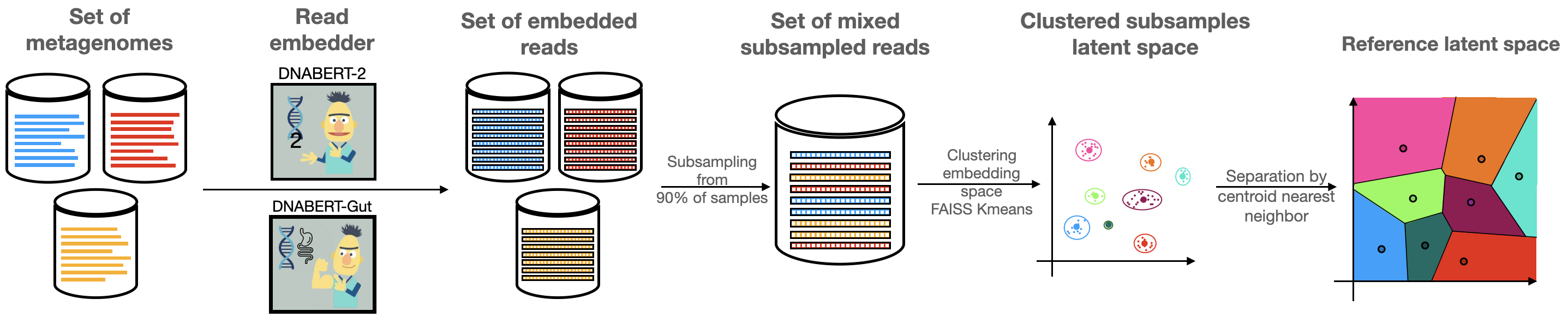}
        \caption{Computing a Reference Latent Space}
    \end{subfigure}
    \hfill
    \begin{subfigure}[b]{\textwidth}
        \centering
        \includegraphics[width=\textwidth]{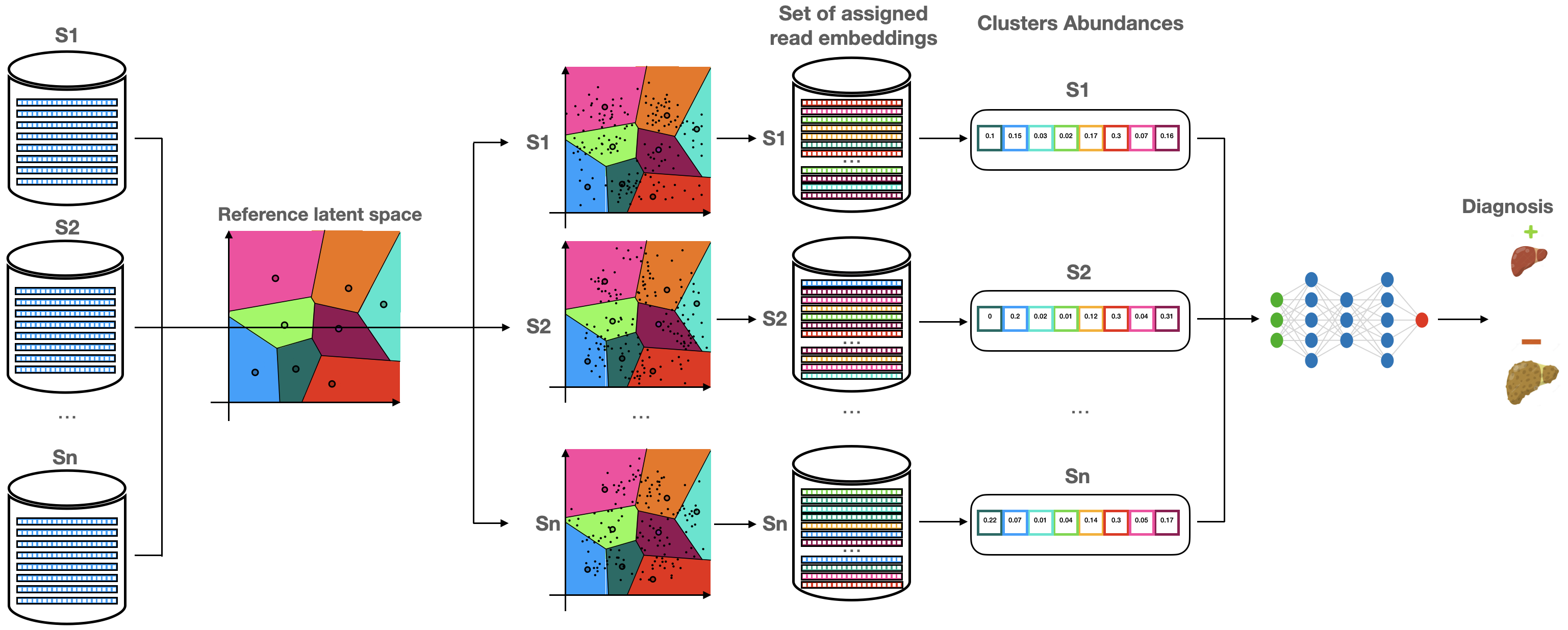}
        \caption{Projecting samples in the reference space}
    \end{subfigure}
    \hfill
   \caption[\textbf{MetagenBERT-Glob: The Global Clustering Architecture.}]{\textbf{MetagenBERT-Glob: The Global Clustering Architecture.} \textit{Panel (a).} For each metagenome, all reads are embedded through the use of a gLLM. A subsample of each embedded metagenome is used to train a KMeans common to all the selected dataset. The centroids of the clusters obtained through K-means are retrieved, effectively partitioning the latent space. \textit{Panel (b).} Every read embedding of each sample is then assigned to its cluster by nearest neighbor search against the centroids, thus creating a new abundance vector based on embeddings rather than species used for classification.}
   \label{fig:archi_clust_global}
 \end{figure}

The entire pipeline is displayed in Figure \ref{fig:archi_clust_global}.

%% file: LaTeX/sections/2.Materials_And_methods/2.2.Datasets.tex
To assess the performance of MetagenBERT in disease classification, we used 5 shotgun metagenomic datasets representing various gut microbiome associated diseases along with control samples that have been used as benchmarks in evaluating many different classification methods : following their use in MetAML \cite{pasolli_machine_2016}, they have also been used in testing DeepMicro \cite{oh_deepmicro_2020}, PopPhy-CNN \cite{reiman_popphy-cnn_2020} or EnseDeepDP \cite{shen_ensdeepdp_2022}.

These datasets represent five diseases : Liver Cirrhosis (LC) \cite{cirr-dataset}, Colorectal Cancer (CRC) \cite{crc-dataset}, Inflammatory Bowel Disease (IBD) \cite{ibd-dataset}, Obesity (Obe) \cite{obe-dataset} and Type 2 Diabetes (T2D) \cite{t2d-dataset}. Noteworthy, one sample is corresponding to one metagenome, hence composed of millions to tens of millions of short reads. The dataset is paired-end reads. Some information can be found in Table \ref{tab:tablePasolli}

\begin{table}[ht]
  \begin{center}
  \resizebox{\columnwidth}{!}{%
\begin{tabular}{l|c|c|c|c}
                & Total Number  & Number of & Number of & Mean Number of reads  \\
                & of Samples & Control Samples & Case Samples & per sample (stddev)  \\
                \hline

Cirrhosis       & 232            & 114                         & 118            & 51.6M (30.9M)                   \\
Type 2 Diabetes & 344           & 174                        & 170            &  40.2M (11.8M)         \\
Inflammatory Bowel Disease       & 110            & 85                         & 25            & 45.2M (18.4M)                   \\
Colorectal Cancer & 121           & 73                        & 48            &  60.0M (25.5M)         \\
Obesity       & 253            & 89                         & 164            & 68.2M (23.2M)                   \\

\end{tabular}%
}
  \end{center}
  \caption[\textbf{Summary of the gut related diseases datasets used for benchmarking our method.}]{\textbf{Summary of the gut related diseases datasets used for benchmarking our approach.} These datasets have been widely used in metagenomics for disase prediction and present the two main difficulties when classifying metagenomic samples : a very high number of sequences and a small number of samples.}
  \label{tab:tablePasolli}
\end{table}

After developing and testing our model on these benchmark reference datasets such as those cited above, we aimed to use it in order to make a more general model by training it with a bigger, more diverse dataset.

To do so, we used shotgun metagenomic data from the Metacardis project. The Metacardis dataset \cite{metacardis-eugeni} \cite{metacardis1}, coming from the eponymous project, is a large project aiming to obtain different types of data (clinical, metagenomic, metabolomic, RNA-Seq) on a large cohort of patients from different European countries (France, Denmark, Germany), especially patients suffering from Cardiometabolic Diseases (CMD). CMD comprise metabolic (obesity, diabetes) and heart diseases subject to evolution in aging populations. 

The dataset is composed of 2022 samples. Out of these 2022 patients, 894 were French and constitute the sub-cohort "French Metacardis", a third of them comprising longitudinal data. The samples are single-end short reads of variable lengths obtained with IonTorrent sequencing method.

For the Gut Microbiome Disease Benchmark datasets, we used the Fastp tool \cite{chen_ultrafast_2023} with default parameters on paired-ends setting to trim and remove  low-quality sequence reads.
Fastp is an ultrafast quality control and preprocessing tool.

For the Metacardis dataset, no additional cleaning steps were applied as sequences were already cleaned for low quality sequences and decontaminated from host sequences..

%% file: LaTeX/sections/2.Materials_And_methods/2.3.Read_embedding.tex
The first step of MetagenBERT involves embedding all reads from the samples in each dataset by using a gLLM.

The metagenomic samples presented in Section \ref{subsec:datasets} are composed of short reads. Hence, we did not need to use a model integrating the latest innovations handling long range dependencies such as EVO. \cite{evo}
Moreover, our final task, phenotype prediction, is performed at the scale of a metagenomic sample and relies on combining the expressive power of a large number of sequences. We therefore decided to use a gLLM that produces general representations and is not fine-tuned in any task.
Therefore, the first model we used for the read embedding step was DNABERT-2 \cite{zhou2024dnabert2}.

DNABERT-2 was trained on a "multi-species dataset" containing the genomes from up to 135 species separated in 7 categories : fungi, protozoa, mammalians, invertebrates, other vertebrates and bacteria, including 32.49B nucleotides to produce embeddings comprised of 768 features (or dimensions) of sequences with a diverse knowledge of DNA structure.

However, DNABERT-2 is trained on a very diverse cohort of organisms. Considering the fact that gut microbial DNA has followed some specific evolutionary patterns, we figured that specializing the model in producing refined gut microbiome DNA representations could prove useful in the context of our experiments.

We therefore developed a homemade variation of DNABERT-2, namely DNABERT-MS.
Using simulated metagenomic reads from 4644 reference genomes from the UHGG catalog \cite{almeida}, we adapted DNABERT-2 for gut microbiome DNA through continuous pre-training with a Masked Language Modeling (MLM) objective (15\% masking). Training ran for three epochs with a batch size of 1,536 over five days on two A100 GPUs, producing the specialized model DNABERT-MS (“Metagenomic Species”).

To compare this model to DNABERT-2, we designed two evaluation tasks.
The intrinsic validation task consists in measuring the cross entropy of the models when reconstructing masked tokens.  10,000 reads per sample from the five benchmark datasets were used. This setup enabled assessment of the model’s representational and compressive capacity on real metagenomic sequences after training on simulated data.

\begin{figure}[!htp]
   \begin{center}
     \setlength{\unitlength}{5mm}
     \includegraphics[width=1\textwidth]{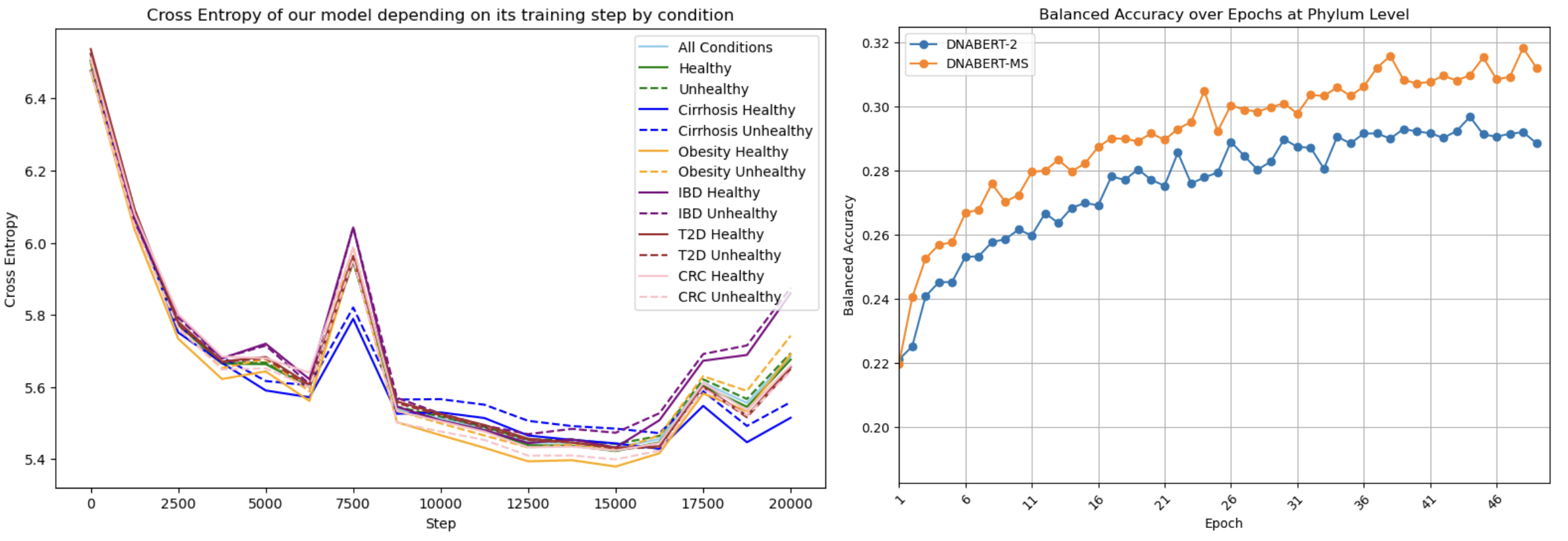}
   \end{center}
   \caption[\textbf{Evaluation of DNABERT-MS against DNABERT-2 with intrinsic and extrinsic validation.}]{\textbf{Evaluation of DNABERT-MS against DNABERT-2 with intrinsic and extrinsic validation.} On the left panel is the cross-entropy of DNABERT-MS for each checkpoint computed on real world evaluation datasets. The cross-entropy was calculated using DNABERT-2 for step 0, then the different checkpoints of DNABERT-MS on benchmark gut-related diseases datasets. We show only the first checkpoints for easy visualization. The cross-entropy is decreasing up to the 15.000 step, then the model appears to diverge due to overfitting. On the right panel is balanced accuracy of Phylum classification of real world sequences, showing consistent amelioration when using DNABERT-MS embeddings.}
   \label{fig:real-data-loss}
 \end{figure}

Training showed rapid early improvement followed by slower convergence, typical of LLM dynamics (Figure \ref{fig:real-data-loss}). Loss decreased from 6.5 to 5.4 before plateauing around step 15,000, suggesting early capture of metagenomic structure but mild overfitting with extended exposure. 

We therefore selected the model checkpoint at step 15,000 as the final DNABERT-MS, achieving the best compression on benchmark datasets. This version is used for subsequent read embedding (alongside DNABERT-2 and DNABERT-S) and for downstream metagenomic classification tasks.

The second task, the extrinsic validation task, conisted in taxonomic classification of 5,000,000 reads from case samples for each benchmark datasets. The species and phyla of origin of these reads were retrieved through Centrifuge \cite{centrifuge}.

Then, we trained a simple classification layer to predict the phylum of origin of each sequence. One model was trained with DNABERT-2 embeddings and one with DNABERT-MS embeddings.

We observe that both models gradually learn to classify metagenomic reads, indicating that the embeddings generated by these models contain useful information for classification, even for a difficult task using such simple models.

DNABERT-MS classification consistently outperforms DNABERT-2 classification, demonstrating that it has indeed learned features specific to this metagenomic dataset. However, the performance gap, while consistent, does not represent an important improvement over DNABERT-2 embeddings.

Overall, combining the insights from both compression and taxonomic classification, we infer that DNABERT-MS has captured specific metagenomic DNA features.

We used both models for read embedding.
These models were used solely in inference mode to generate embeddings without any additional training. For each sequence, token embeddings are pooled using either mean or max pooling. Preliminary tests indicated that max pooling produced lower-quality embeddings, leading us to adopt mean pooling for all subsequent experiments. Additionally, we retained indexes linking each sequence’s position in the FASTA file to its embedding, ensuring traceability. Details of this step are provided in Supplementary Material.

Given the large dataset size, we utilized the Jean Zay supercomputer. Embeddings were processed in batches of 40,000 sequences, each requiring approximately 14 seconds on a single A100 GPU and 118 MB of storage. For the full embedding of the cirrhosis and T2D datasets, we used 96 GPUs.

Embedding a full 40M-sequence metagenome on a single GPU would need 3 hours and 50 minutes. By leveraging high-performance parallelization across multiple GPUs, this time is substantially reduced; theoretically, embedding all 232 samples of the cirrhosis dataset could be completed in ~9 hours and 40 minutes. However, practical execution introduces overhead due to inter-node communication, model/sample loading, imperfect batch distribution, and dependence on the slowest GPU. In our experiments, embedding the entire cirrhosis dataset required approximately 12 hours, while the T2D dataset required 14 hours, corresponding to an average of approximately 3 minutes per sample. Although this runtime is low, it presupposes access to extensive computing resources.
We show in Table \ref{tab:comp-metagenbert} the time consumed for embedding each sample with gLLM models and in Table \ref{tab:comp-sota} the time used to obtain a taxonomic profile with various State of the Art methods

\begin{table}[]
\resizebox{\columnwidth}{!}{%
\begin{tabular}{l|l|l|l|l|l|l|l}
         & Cirrhosis & T2D      & IBD      & CRC      & Obesity  & Metacardis & Patient   \\
         \hline
Temps    & 12h07min  & 14h00min & 5h01min  & 7h21min  & 17h28min & 47h43min   & 3min02sec \\
\hline
Stockage & 35.614 T  & 40.794 T & 14.667 T & 21.417 T & 50.901 T & 138.982 T  & 147.5G   
\end{tabular}%
}

\caption{\textbf{Time and storage required for the embedding of benchmark datasets, metacardis dataset and a typical patient sample using MetagenBERT with 96 A100 GPUs.} The high parallelization of the embedding step makes it feasible in an amount of time of the order of the minute, but the storage needed for embeddings is still counted in tens of Terabytes.}
\label{tab:comp-metagenbert}
\end{table}

\begin{table}[]
\resizebox{\columnwidth}{!}{%
\begin{tabular}{l|l|l|l|l|l|l|l}
          & Cirrhosis     & T2D       & IBD        & CRC           & Obesity       & Metacardis    & Patient      \\
          \hline
Kraken2   & $\sim$7h      & $\sim$8h  & $\sim$2h30 & $\sim$4h      & $\sim$9h30min & $\sim$20h     & $\sim$1min30 \\
\hline
Sylph     & $\sim$2h30min & $\sim$3h  & $\sim$1h   & $\sim$1h30min & $\sim$4h      & $\sim$8h30min & $\sim$30sec  \\
\hline
MetaPhlAn & $\sim$77h     & $\sim$90h & $\sim$30h  & $\sim$44h     & $\sim$105h    & $\sim$220h    & $\sim$20min 
\end{tabular}%
}
\caption{\textbf{Time required to compute the taxonomic profiles of benchmark datasets, metacardis dataset and a typical patient sample using reference methods Kraken2 \cite{kraken2}, Sylph \cite{sylph} and MetaPhlAn \cite{blanco-miguez_extending_2022} with 36 CPUs.} The computation of a taxonomic profile using the fastest method, Sylph, is faster than MetagenBERT and comparable when using Kraken2, leading us to downsample the number of reads used for metagenome embedding}
\label{tab:comp-sota}
\end{table}

The times needed for both types of methods are comparable, but embedding through a gLLM requires substantially larger computational resources, making this method hardly scalable.
Moreover, storing all the embeddings requires tens of Terabytes, representing a huge challenge.
For these reasons, we decided to use only 10\% of the reads for each sample. The time required for embedding decreases linearly and can therefore be divided by ten. We justify this choice in Section \ref{subsec:undersampling}.

%% file: LaTeX/sections/2.Materials_And_methods/2.4.Metagenome_Embedding.tex
After the read embedding step, each metagenome is represented by an unordered set of millions of embedding vectors of dimension 768, each corresponding to a sequence of the original metagenome. 
Our objective is now to represent this large point cloud with a more simple embedding. We needed this embedding to be both simple and interpretable across samples. When using species abundances, we can compare various samples by looking for precise species. We sought to produce an embedding that could allow such a comparison to mitigate the black box nature of Transformer embeddings.

To do so, we decided to group the reads of each sample into clusters.
In order to create a consistent representation among clusters, we trained a "global clustering" method. We selected 90\% of the samples for each cohort and randomly drew 250 000 reads from each of these samples and trained a clustering algorithm on this set of reads.
Given the massive scale and high dimensionality of our metagenomes, we opted for K-Means, as it offers superior scalability and acceptable robustness in this context, with the large number of points ensuring robustness to outliers. We used the FAISS GPU-optimized K-Means algorithm for efficiency \cite{johnsonbillion,douze2025faisslibrary}. We tested different numbers of clusters : 128, 512, 2048 and 8192. 

Once the global K-Means model is trained, each sample is processed by assigning all reads to their nearest cluster centroid. This produces, for each read, a cluster label and, for each sample, a distribution of reads across clusters. MetagenBERT-Glob represents each sample as a single abundance vector, reflecting the proportion of reads assigned to each global cluster.

To evaluate the effectiveness of our embedding-based abundance representation relative to conventional species-based abundance, we retrieved species-abundance vectors for each sample and trained disease prediction models under three configurations: (1) using species abundance alone, (2) using clustering-based abundance vectors alone (for varying numbers of clusters), and (3) using a concatenation of both representations. This experimental design not only provides a direct comparison with the conventional baseline but also allows us to assess whether the two approaches capture complementary information from the metagenome.

For classification, we employed a LASSO model \cite{linear-rregression}. This choice contrasts with state-of-the-art methods, which often incorporate additional sources of information (e.g., phylogenetic structure in PopPhyCNN \cite{reiman_popphy-cnn_2020}, multimodal data in MML4Microbiome \cite{lee_multimodal_2022}) or rely on more complex learning strategies, such as convolutional neural networks or ensemble approaches (e.g., EnsDeepDP \cite{shen_ensdeepdp_2022}).

To facilitate its future use in clinical applications, our goal is to develop a model that can generalize across a broad range of phenotypes. 

The central challenge is therefore to define a partition of the embedding space that is sufficiently comprehensive to enable reuse across new, unseen and heterogeneous phenotypes without retraining the embedding space, hoping to construct a "foundation" model for metagenome embedding.

Given its large sample size and extensive phenotypic diversity, the MetaCardis dataset represents an ideal candidate for such pretraining. To this end, we trained new K-means models from 100,000 reads from the 894 French MetaCardis samples using both DNABERT-2 and DNABERT-MS.

Subsequently, for each gut benchmark dataset, we applied the same assignment strategy as in the Global Clustering approach, except that cluster centroids were derived from the MetaCardis-trained K-means rather than dataset-specific clustering. The resulting vectors were then used to train a LASSO classifier for each dataset, and classification scores were computed accordingly. In summary, this procedure adapts the MetagenBERT-Glob framework by replacing dataset-specific clusters with those inferred from the MetaCardis data, with the goal of producing more generalizable metagenome embeddings. We refer to this approach as \textbf{MetagenBERT-Glob-MCardis}.

Due to the very high number of reads contained in a metagenome and the accompanying computational difficulties developed in Section \ref{subsec:read}, we applied both methods by using only 10\% of the reads for each metagenome (be it for selecting the embeddings used for clustering training or for assignment to a cluster), a choice we will justify in Section \ref{subsec:undersampling}.

%% file: LaTeX/sections/3.Results/3.Results.tex
\subsection{Disease Prediction}
\label{subsec:pred}
\input{LaTeX/sections/3.Results/3.1.Disease_Prediction}

\subsection{Undersampling Reads inside Samples}
\label{subsec:undersampling}
\input{LaTeX/sections/3.Results/3.2.Undersampling_results}

\subsection{Clusters Analysis}
\label{subsec:metagenome}
\input{LaTeX/sections/3.Results/3.4.Clusters_Analysis}

%% file: LaTeX/sections/3.Results/3.1.Disease_Prediction.tex
We compare our method to state-of-the-art approaches, including a LASSO model trained only on species abundances (MetagenBERT-Abu). We evaluated our models based on cluster abundances alone (named MetagenBERT-N-Glob-XX, where "N" denotes the embedder—either "2" or "MS"—and "XX" represents the number of clusters when we use cluster abundance alone) and when concatenating both types of abundances (MetagenBERT-N-Glob-XX-Abu).

First, when comparing MetagenBERT-N-Glob-XX with MetagenBERT-Abu, which uses only species abundance, we observe that the former consistently competes with or even outperforms the latter (notably on T2D, IBD, and CRC), indicating that this model effectively captures relevant dynamics in each dataset that are at least as informative as species abundances.

Furthermore, concatenating both sources of information generally improves performance, supporting the notion that species abundances and cluster abundances encode complementary, rather than redundant, information. Two exceptions are observed in the MetagenBERT-MS-Glob case for IBD and T2D, where adding species abundance appears to slightly obscure the signal. In particular, for IBD, the model achieves strong results without species abundance, with a slight decrease when it is added. 
Unexpectedly, regarding type 2 diabetes, adding species abundances to MetagenBERT-MS-Glob degrades performance, whereas it slightly improves it with MetagenBERT-2-Glob.

Compared to state-of-the-art methods, our approaches remain competitive across all tasks, falling short only relative to EnsDeepDP on CRC and Obesity, while outperforming other models on T2D and IBD.

With the exception of the Obesity dataset, increasing the number of clusters generally correlates with improved performance, although the effect is less pronounced than in MetagenBERT-Sub and exhibits some outliers (notably IBD). This suggests that higher granularity can enhance accuracy, but the improvement is moderate.

Finally, across datasets, DNABERT-2 and DNABERT-MS yield very similar classification performance. This suggests that phenotype discrimination primarily arises from sample-level structure in the embedding space and inter-sample variability rather than from fine-grained differences in read-level embeddings, which is coherent with the fact that MetagenBERT-Glob relies on the repartition of a large number of sequences rather than on a few specific discriminating reads. This observation is consistent with the fact that DNABERT-MS is trained to better represent microbiome DNA, but not necessarily to differentiate sequences associated with specific phenotypes.

We propose that developing a model specifically designed to optimize phenotypic differentiation could be valuable for metagenome embedding. 

\begin{figure}[htbp]
    \centering
    \begin{subfigure}[b]{0.49\textwidth}
        \centering
        \includegraphics[width=\textwidth]{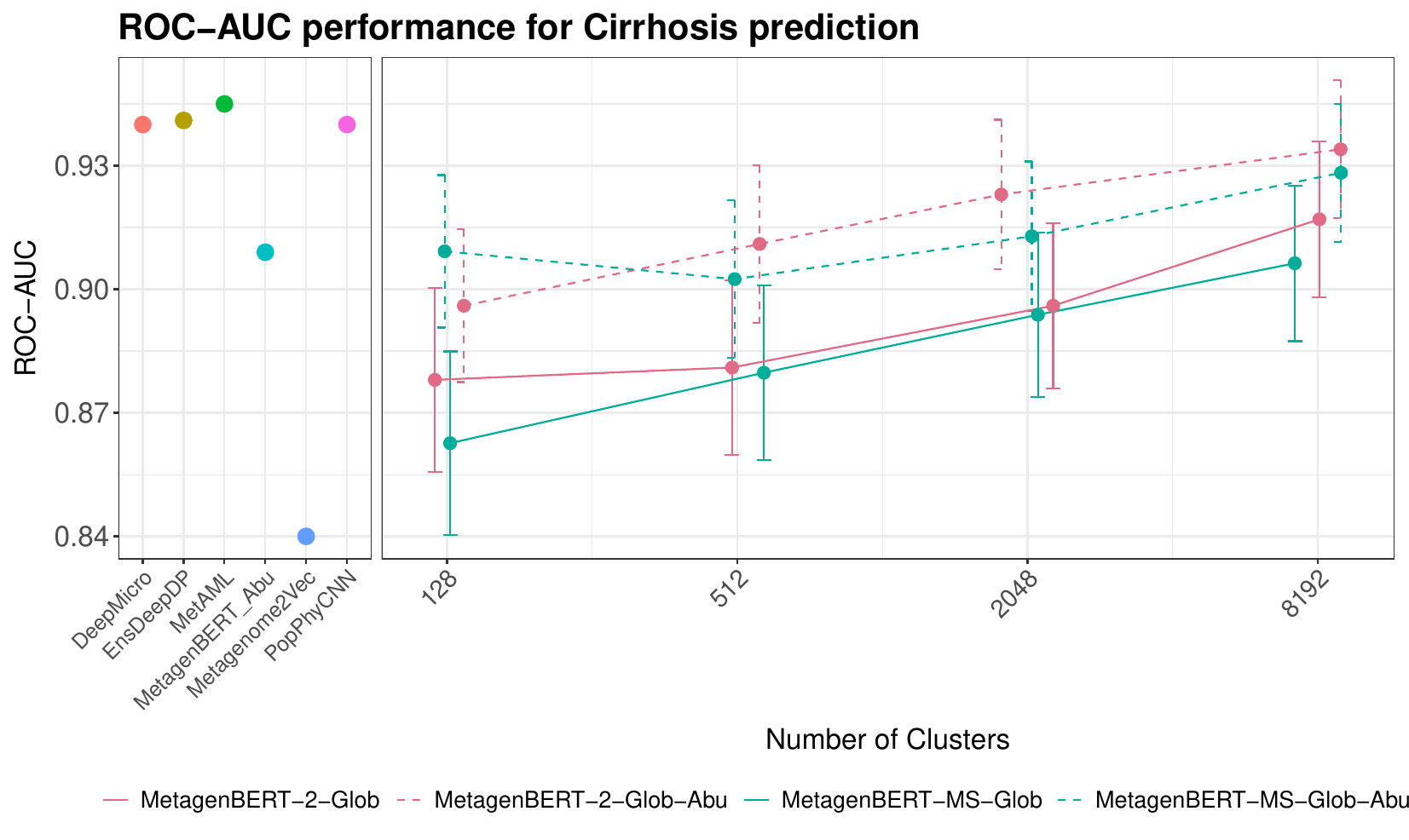}
        \caption{MetagenBERT-Glob performance on Cirr dataset}
    \end{subfigure}
    \hfill
    \begin{subfigure}[b]{0.49\textwidth}
        \centering
        \includegraphics[width=\textwidth]{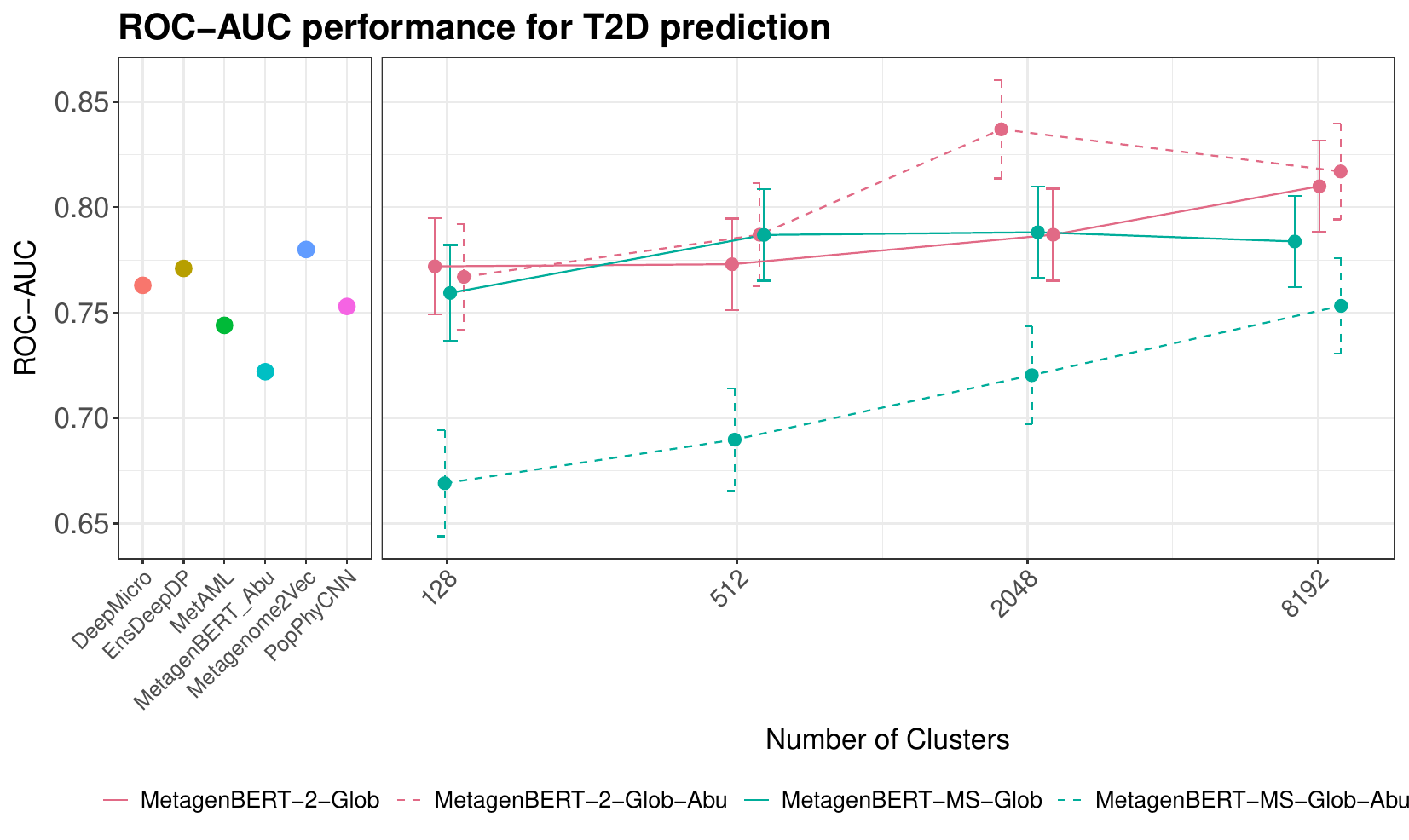}
        \caption{MetagenBERT-Glob performance on T2D dataset}
    \end{subfigure}
    \hfill
    \begin{subfigure}[b]{0.49\textwidth}
        \centering
        \includegraphics[width=\textwidth]{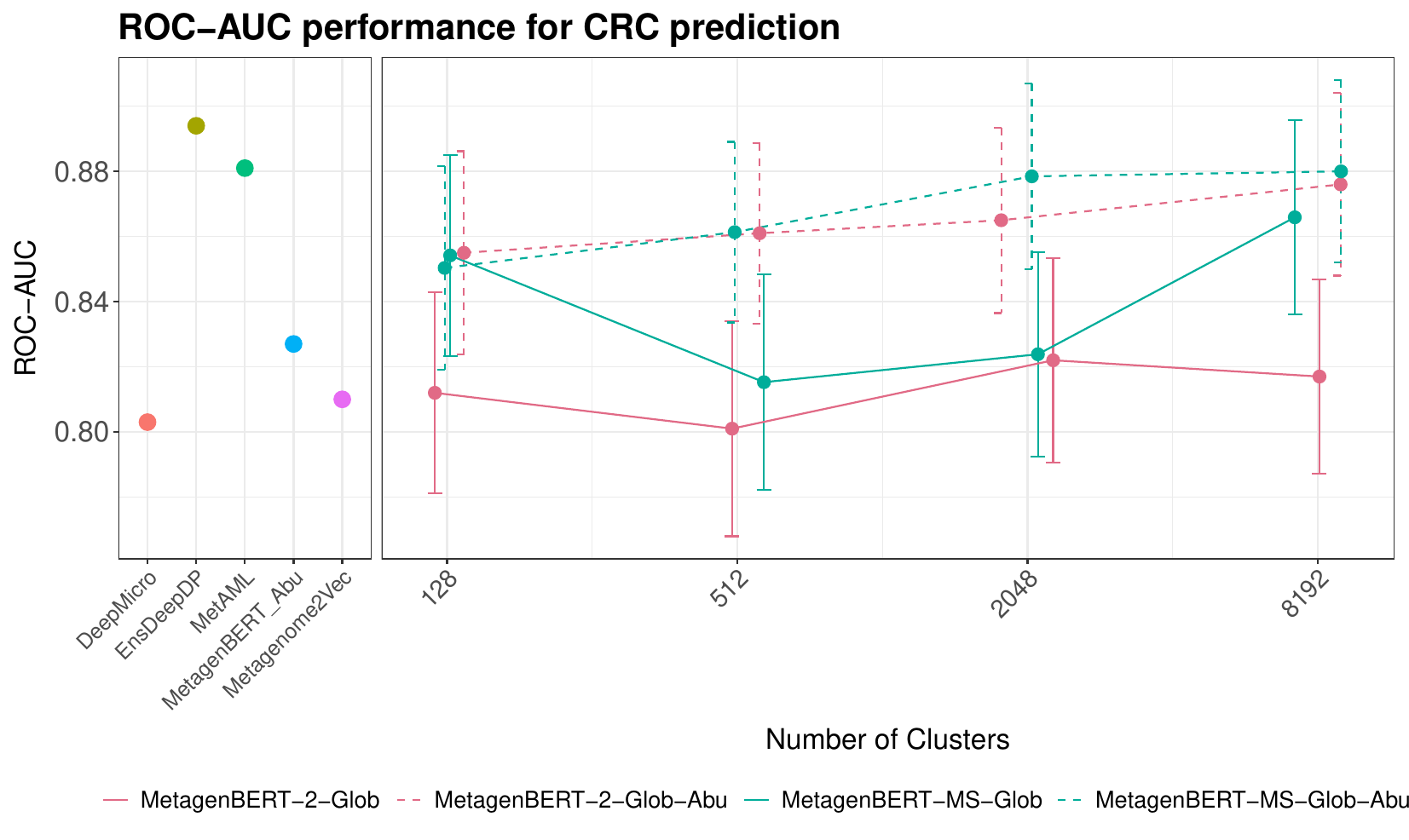}
        \caption{MetagenBERT-Glob performance on CRC dataset}
    \end{subfigure}
    \hfill
    \begin{subfigure}[b]{0.49\textwidth}
        \centering
        \includegraphics[width=\textwidth]{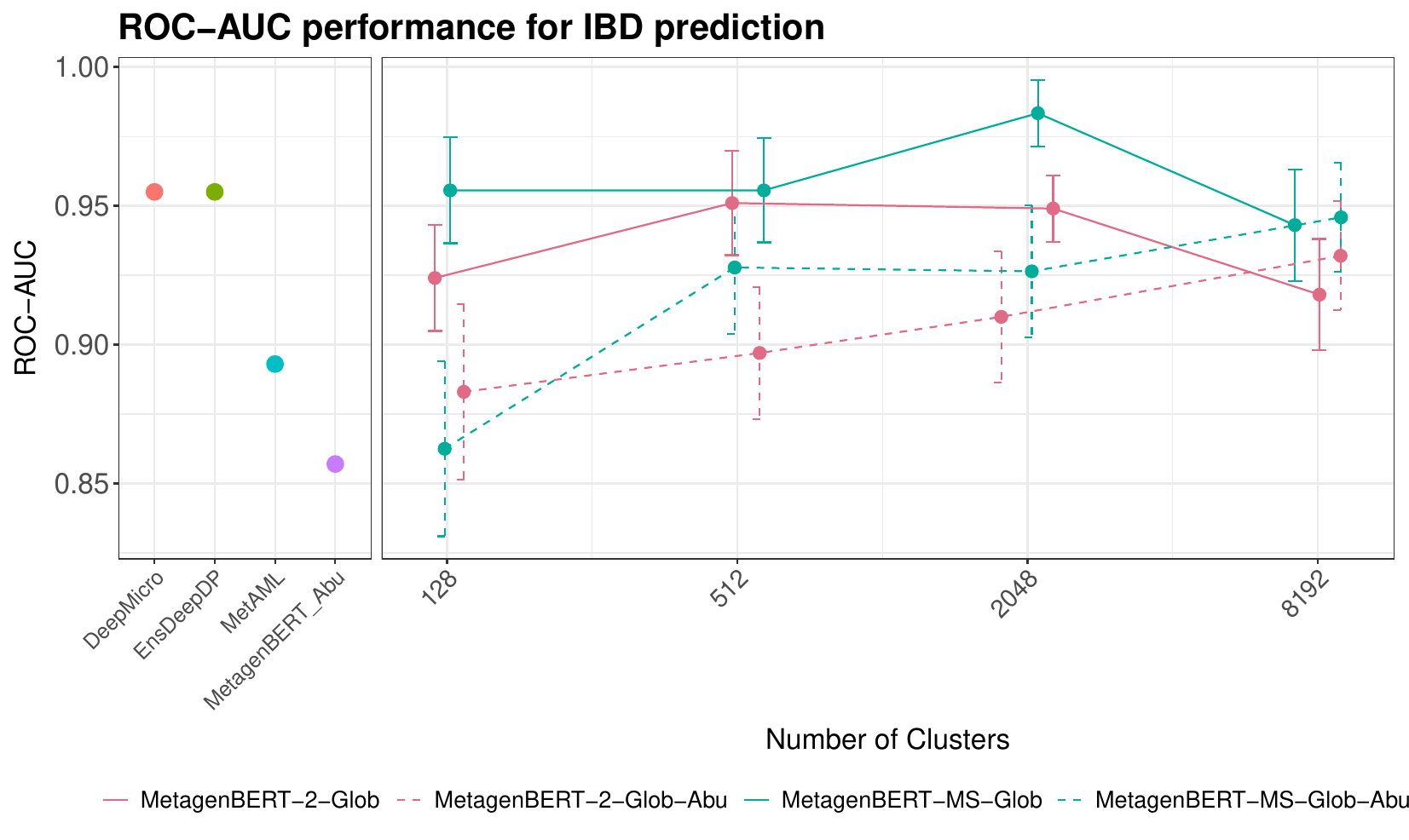}
        \caption{MetagenBERT-Glob performance on IBD dataset}
    \end{subfigure}
    \hfill
    \begin{subfigure}[b]{0.49\textwidth}
        \centering
        \includegraphics[width=\textwidth]{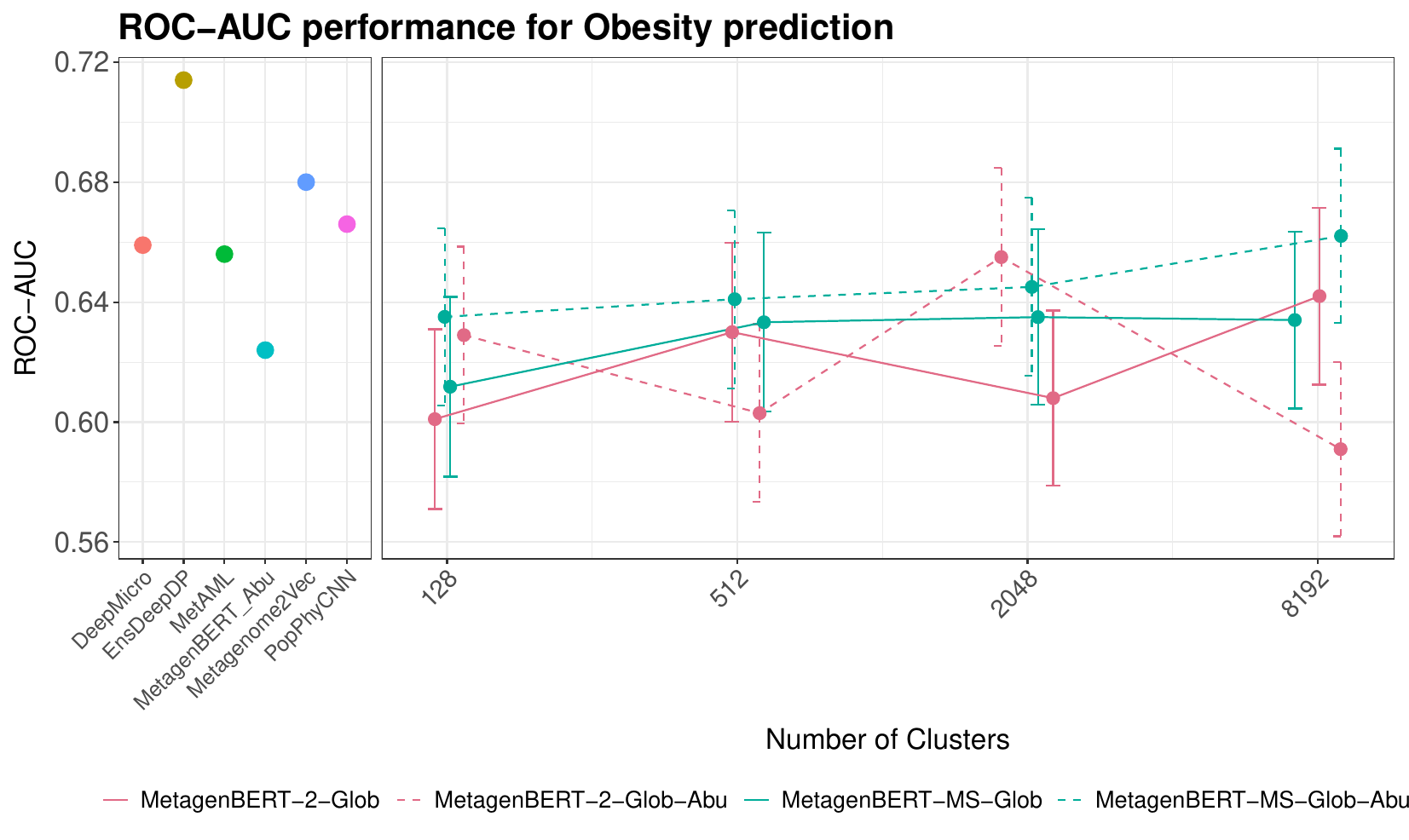}
        \caption{MetagenBERT-Glob performance on Obesity dataset}
    \end{subfigure}
    \hfill
    \caption[\textbf{Performances of MetagenBERT-Glob on 5 benchmark datasets with various numbers of vectors and embedders}]{\textbf{Performances of MetagenBERT-Glob on 5 benchmark datasets with various numbers of clusters and embedders}. MetagenBERT-Glob competes with a range of state-of-the-art methods across all datasets and tends to outperform them on T2D. In most configurations, although not universally, increasing the number of clusters generally leads to higher AUC. “MetagenBERT-Abu” refers to the performance of a LASSO model trained such as the one we used with MetagenBERT embeddings, but trained solely on species abundances. Across all five datasets, MetagenBERT-Glob either matches or surpasses the performance of the species-abundance-only model. When species abundances are concatenated with cluster abundances, performance generally improves, supporting the hypothesis that the two types of information are complementary. An exception is observed in the IBD dataset, where MetagenBERT-Glob performs better without the addition of species abundance. It is difficult to conclude that DNABERT-MS consistently outperforms DNABERT-2, as both embedders produce very similar results, with the notable exception of T2D, where MetagenBERT-MS performs unexpectedly poorly. The margins reported correspond to standard errors.}
    \label{fig:glob_results}
\end{figure}

\begin{table}[]
\resizebox{\columnwidth}{!}{%
\begin{tabular}{l||lllll}
                        & Cirrhosis & T2D   & Obesity & IBD   & CRC   \\
                        \hline
                        \hline
MetAML \cite{pasolli_machine_2016}                  & \textbf{0.946}     & 0.745 & 0.656   & 0.893 & 0.881 \\
PopPhyCNN \cite{reiman_popphy-cnn_2020}              & \textbf{0.946}     & 0.69  & 0.666   &       &       \\
DeepMicro \cite{oh_deepmicro_2020}              & 0.940     & 0.763 & 0.659   & \textbf{0.955} & 0.803 \\
EnsDeepDP \cite{shen_ensdeepdp_2022}              & 0.941     & 0.771 & \textbf{0.714}   & \textbf{0.955} & \textbf{0.894} \\
Metagenome2Vec \cite{queyrel_towards_2020}         & 0.84      & \textbf{0.78}  & 0.68    &       & 0.81  \\
\Xhline{3\arrayrulewidth}
LASSO         & 0.909     & 0.722 & 0.624   & 0.857 & 0.827 \\
\Xhline{3\arrayrulewidth}
MetagenBERT-2-Glob-8192        & \textbf{0.917}     & \textbf{0.810} & \textbf{0.642}   & 0.918 & 0.817 \\
MetagenBERT-MS-Glob-8192     & 0.906     & 0.784 & 0.634   & \textbf{0.943} & \textbf{0.866} \\
\Xhline{3\arrayrulewidth}
MetagenBERT-2-Glob-8192-Abu    & \textbf{0.934}     & \textbf{0.817} & 0.591   & 0.932 & 0.876 \\
MetagenBERT-MS-Glob-8192-Abu & 0.928     & 0.753 & \textbf{0.662}   & \textbf{0.945} & \textbf{0.880}
\end{tabular}%
}
\caption[\textbf{AUC Performances of MetagenBERT-Glob}]{\textbf{AUC Performances of MetagenBERT-Glob.} This table compares our variations of MetagenBERT-Glob to State of The Art methods. The following model use MetagenBERT clusters abundances for 8192 clusters while MetagenBERT-N-Glob-8192-Abu uses the concatenation of both species abundance and clusters abundance. In Supplementary Material can be seen the results for other numbers of clusters resulting in some cases in better results, outperforming State of the Art on IBD dataset.}
\label{tab:tab-glob-perf}
\end{table}

\begin{figure}[!htp]
   \begin{center}
    \centering
    \begin{subfigure}[b]{0.49\textwidth}
        \centering
        \includegraphics[width=\textwidth]{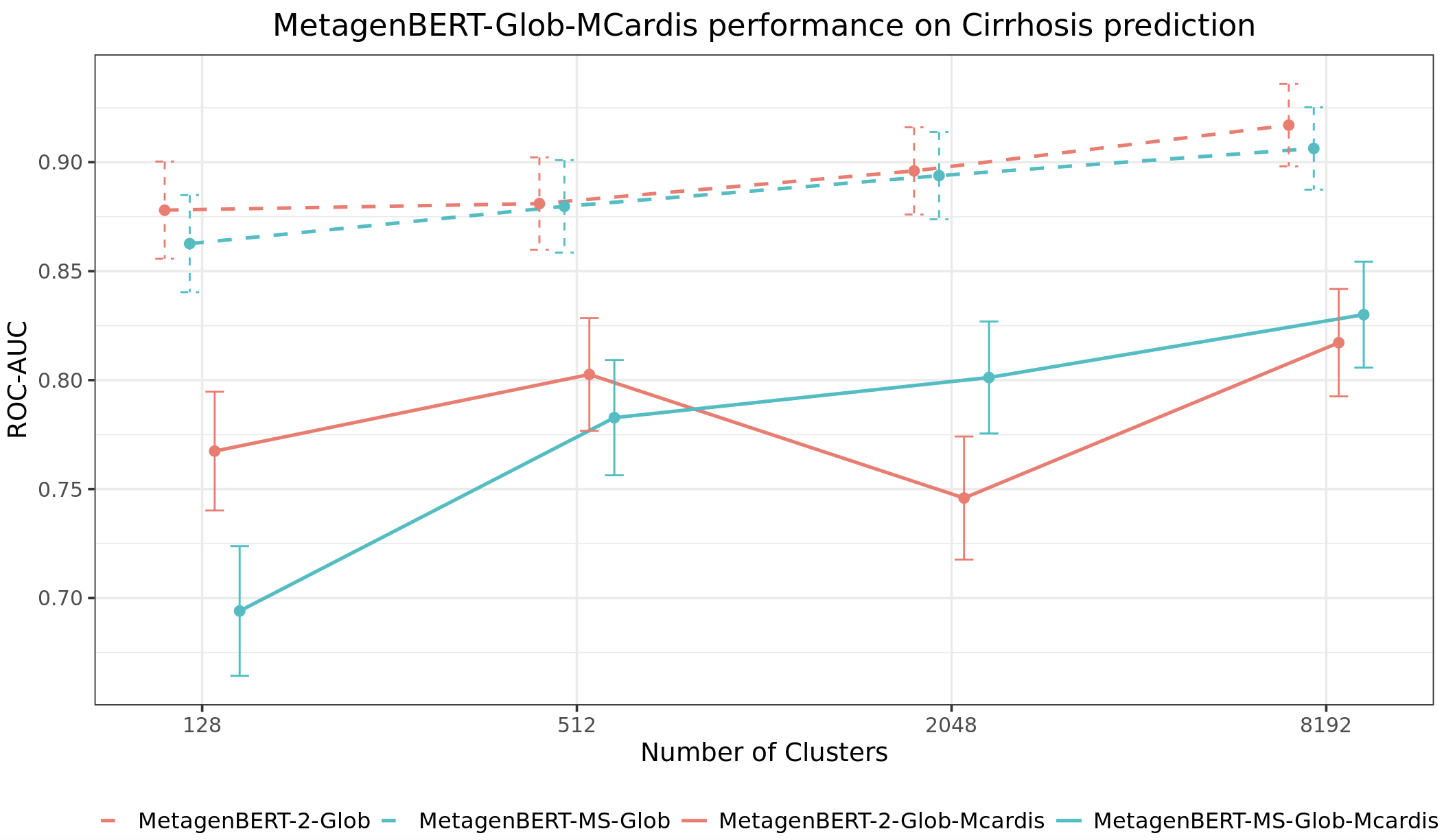}
        \caption{MetagenBERT-Glob and MetagenBERT-Glob-Mcardis performances on Cirrhosis dataset}
    \end{subfigure}
    \hfill
    \begin{subfigure}[b]{0.49\textwidth}
        \centering
        \includegraphics[width=\textwidth]{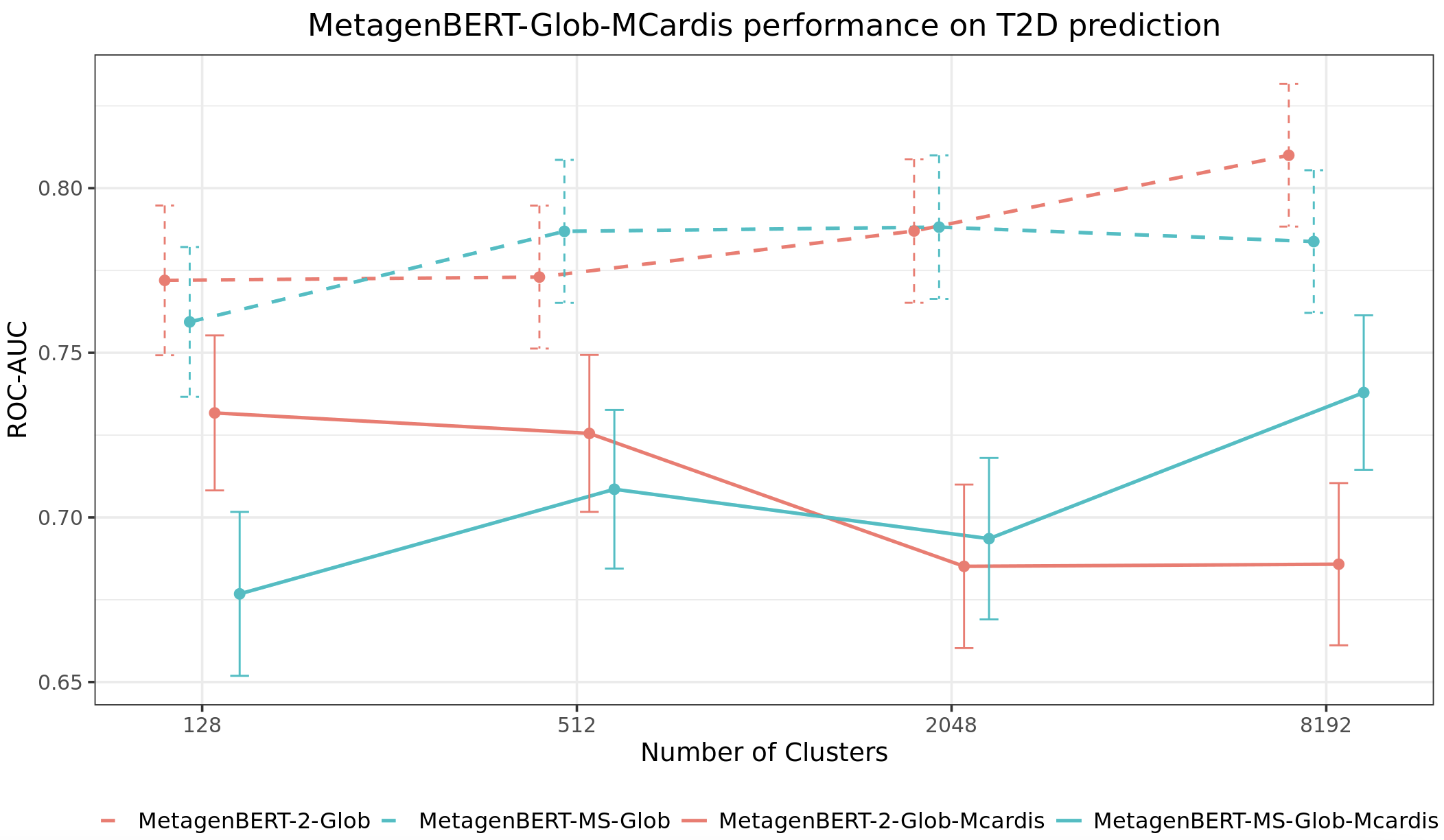}
        \caption{MetagenBERT-Glob and MetagenBERT-Glob-Mcardis performances on T2D dataset}
    \end{subfigure}
    \hfill
    \begin{subfigure}[b]{0.49\textwidth}
        \centering
        \includegraphics[width=\textwidth]{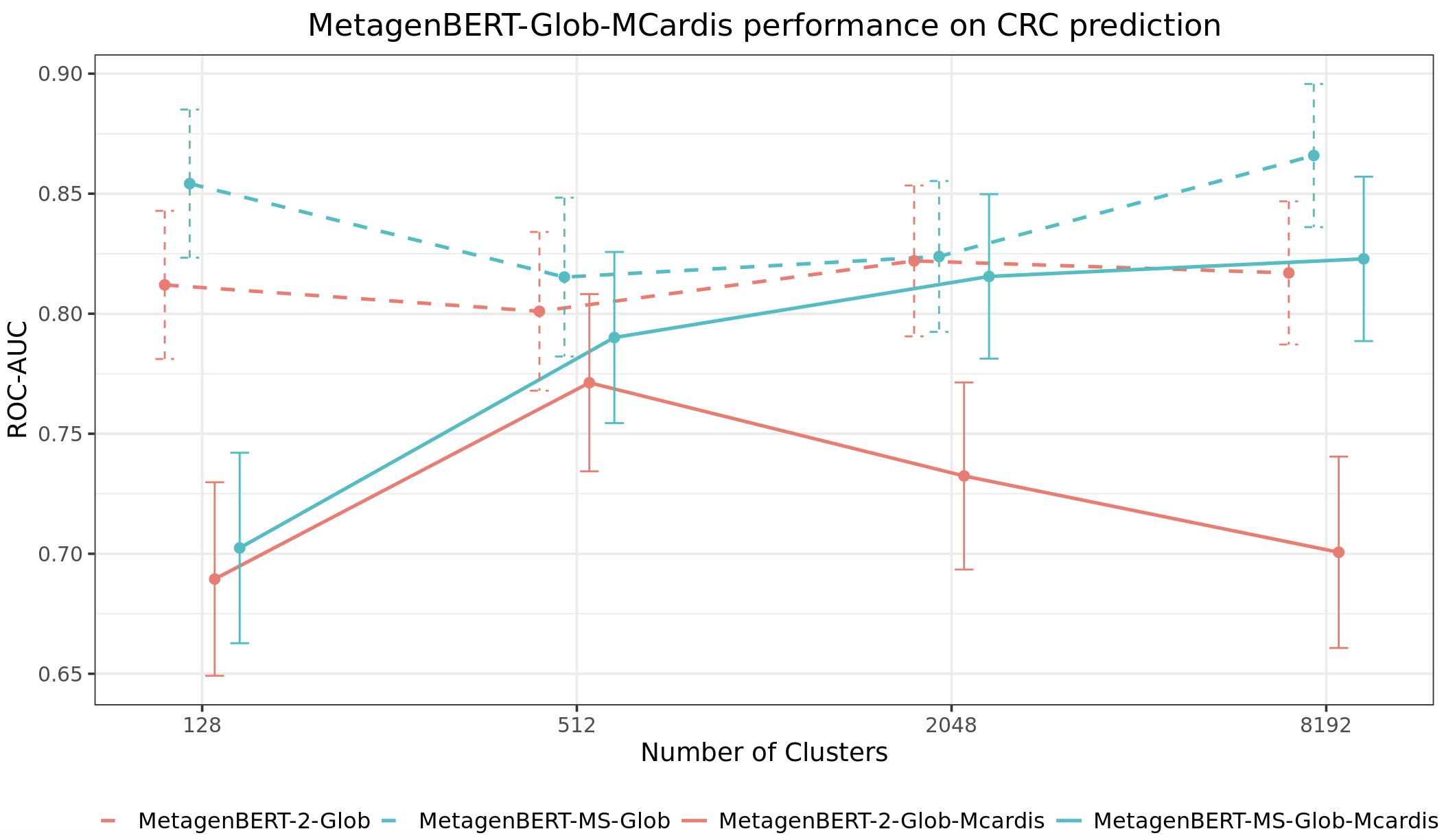}
        \caption{MetagenBERT-Glob and MetagenBERT-Glob-Mcardis performances on CRC dataset}
    \end{subfigure}
    \hfill
    \begin{subfigure}[b]{0.49\textwidth}
        \centering
        \includegraphics[width=\textwidth]{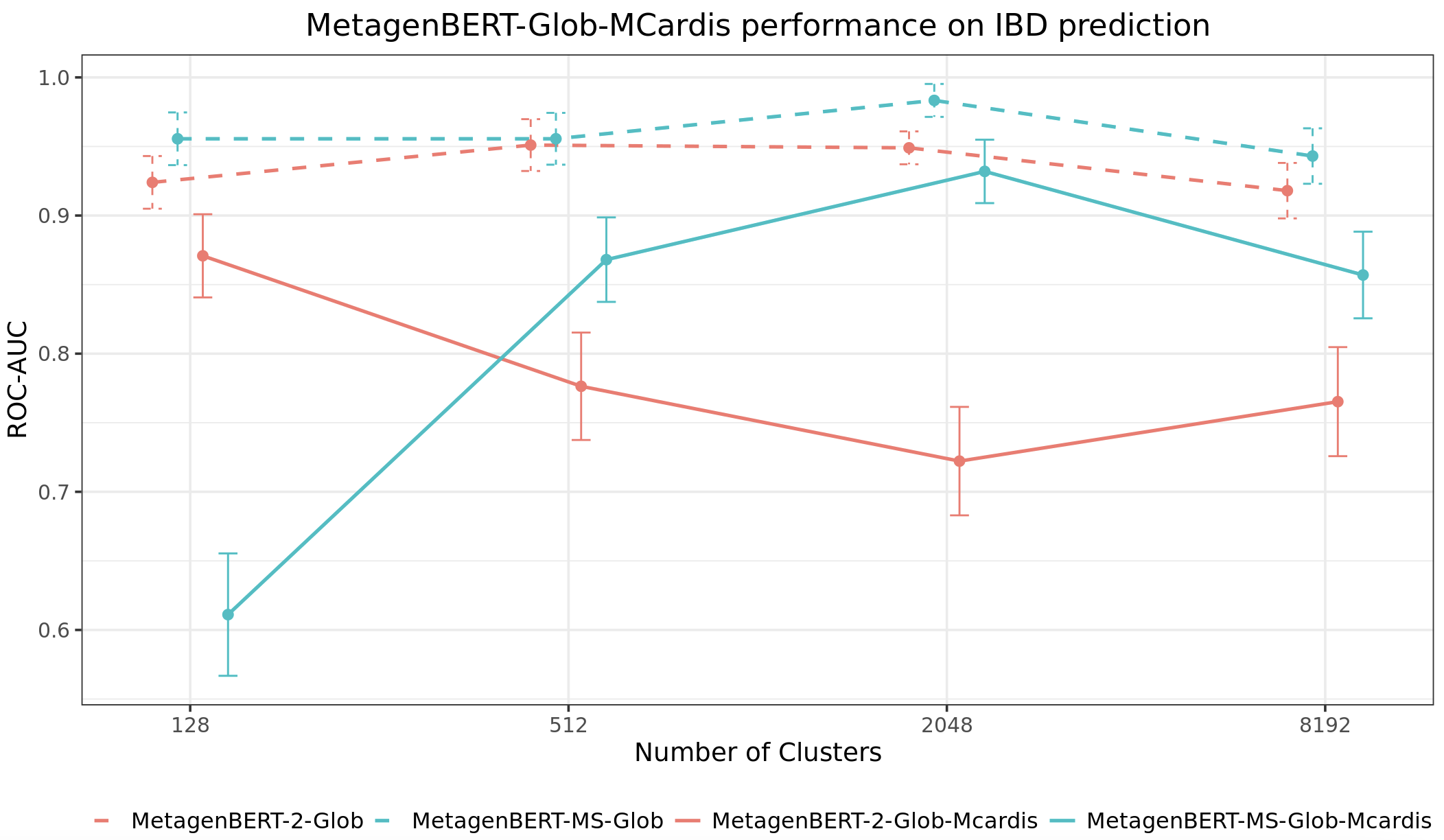}
        \caption{MetagenBERT-Glob and MetagenBERT-Glob-Mcardis performances on IBD dataset}
    \end{subfigure}
    \hfill
    \begin{subfigure}[b]{0.49\textwidth}
        \centering
        \includegraphics[width=\textwidth]{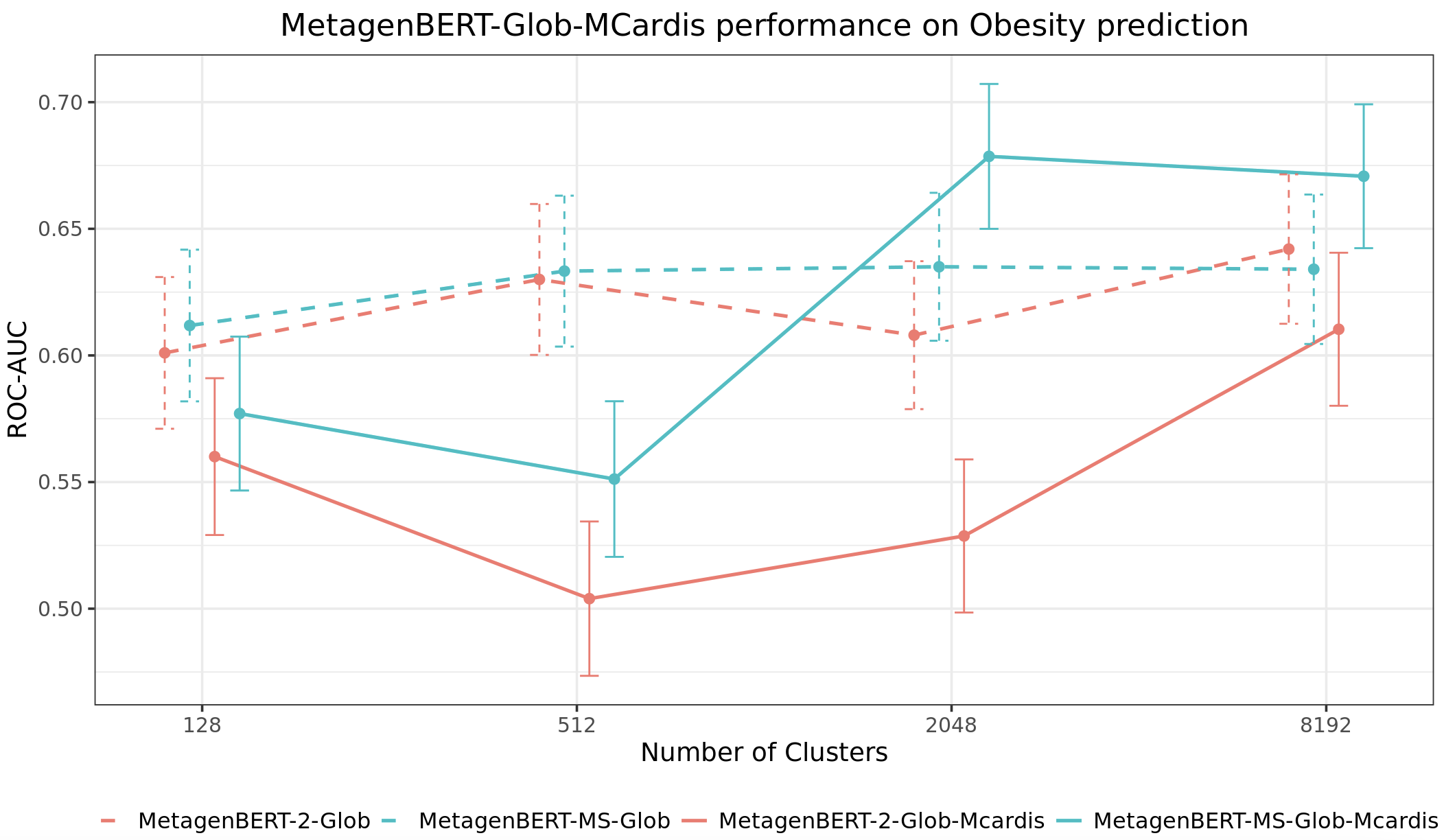}
        \caption{MetagenBERT-Glob and MetagenBERT-Glob-Mcardis performances on Obesity dataset}
    \end{subfigure}
    \hfill
    \end{center}
   \caption[\textbf{Comparison of Prediction performance when using MetagenBERT-Glob trained directly on each benchmark dataset or when using cluster abundance obtained through Metacardis trained K-means.}]{\textbf{Comparison of Prediction performance when using MetagenBERT-Glob trained directly on each benchmark dataset or when using cluster abundance obtained through Metacardis trained K-means.} The figure shows that, even if the performances are globally lower than when clustering on the dataset of origin, this dataset efficiently captures important dynamics even with Metacardis K-means alone.}
   \label{fig:foundation_results}
 \end{figure}

Following these experiments, we applied MetagenBERT-Glob-Mcardis with the same configurations. The results, presented in Figure \ref{fig:foundation_results} and Table \ref{tab:tab-assign-perf}, indicate that although performance decreases relative to dataset-specific clustering, the pretrained MetaCardis clusters retain a substantial amount of predictive signal. This is particularly notable in the Cirrhosis and IBD datasets, where the MetaCardis-trained K-means model, despite never having encountered samples with these phenotypes, achieved strong classification performance. Since MetaCardis primarily focuses on cardiometabolic conditions, it is also encouraging that, with a large number of clusters and DNABERT-MS embeddings, obesity prediction outperformed results obtained using dataset-specific clustering. These findings provide promising evidence for the feasibility of developing a foundation model for metagenomics, suggesting that a sufficiently diverse pretraining corpus can support the embedding of novel and heterogeneous phenotypes.

However, several points warrant further consideration. Results exhibit greater variability than expected, with relatively high standard deviations, suggesting some degree of instability in predictions. Another unexpected observation is the inconsistent behavior between DNABERT-2 and DNABERT-MS embeddings, which sometimes display opposite performance trends. For instance, in IBD prediction, DNABERT-2 performs poorly except in the 128-cluster configuration, whereas DNABERT-MS achieves high AUC values in all but that configuration. Similarly, for CRC, DNABERT-2 fails to reproduce the performance improvements observed with DNABERT-MS as the number of clusters increases. While this instability may partly stem from the limited size of these datasets, it also suggests a strong dependence on the clustering structure and its interaction with the underlying embedding space.

\begin{table}[]
\resizebox{\columnwidth}{!}{%
\begin{tabular}{l||lllll}
                              & Cirrhosis & T2D   & Obesity & IBD   & CRC   \\
\hline
\hline
MetagenBERT-2-Glob-128               & 0.878     & 0.772 & 0.601   & 0.924 & 0.812 \\
MetagenBERT-MS-Glob-128               & 0.863     & 0.760 &  0.612  & 0.956 & 0.854 \\
\hline
MetagenBERT-2-Glob-Mcardis-128       & 0.767     & 0.732 & 0.560   & 0.871 & 0.689 \\
MetagenBERT-MS-Glob-Mcardis-128  & 0.694     & 0.677 & 0.577   & 0.611 & 0.702 \\
\Xhline{3\arrayrulewidth}
MetagenBERT-2-Glob-512               & 0.881     & 0.773 & 0.630   & 0.951 & 0.801 \\
MetagenBERT-MS-Glob-512               & 0.880     & 0.787 &  0.633  & 0.956 & 0.815 \\
\hline
MetagenBERT-2-Glob-Mcardis-512       & 0.803     & 0.726 & 0.504   & 0.776 & 0.771 \\
MetagenBERT-MS-Glob-Mcardis-512  & 0.783     & 0.709 & 0.551   & 0.868 & 0.790 \\
\Xhline{3\arrayrulewidth}
MetagenBERT-2-Glob-2048              & 0.896     & 0.787 & 0.608   & 0.949 & 0.822 \\
MetagenBERT-MS-Glob-2048              & 0.894     & 0.788 & 0.635   & \textbf{0.983} & 0.824 \\
\hline
MetagenBERT-2-Glob-Mcardis-2048     & 0.813     & 0.685 & 0.529   & 0.722 & 0.732 \\
MetagenBERT-MS-Glob-Mcardis-2048 & 0.801     & 0.694 & \textbf{0.678}   & \textbf{0.932} & 0.816 \\
\Xhline{3\arrayrulewidth}
MetagenBERT-2-Glob-8192              & \textbf{0.917}     & \textbf{0.810} & \textbf{0.642}   & 0.918 & 0.817 \\
MetagenBERT-MS-Glob-8192              &   0.906   & 0.784 &  0.634  & 0.943 & \textbf{0.867} \\
\hline
MetagenBERT-2-Glob-Mcardis-8192      & 0.817     & 0.686 & 0.610   & 0.765 & 0.701 \\
MetagenBERT-MS-Glob-Mcardis-8192 & \textbf{0.830}     & \textbf{0.738} & 0.671   & 0.857 & \textbf{0.823}
\end{tabular}%
}
\caption[\textbf{AUC performance comparison between MetagenBERT-Glob and MetagenBERT-Glob-Mcardis.}]{\textbf{AUC performance comparison between MetagenBERT-Glob and MetagenBERT-Glob-Mcardis.} As stated before, we see the performances are mostly lower when using Metacardis to train clustering rather than using the dataset itself. Nonetheless, enough features are captured to allow classification, paving the way for a general model able to return embeddings from samples of various sources.}
\label{tab:tab-assign-perf}
\end{table}

%% file: LaTeX/sections/3.Results/3.2.Undersampling_results.tex
As stated before, To further investigate the limits of this approach, we performed global clustering using progressively smaller fractions of reads : 5\%, 3\%, 1\%, 0.5\%, 0.2\%, and 0.1\%, on the cirrhosis dataset. Considering that each sample contains on the order of 10,000,000 reads, the lowest configurations correspond to approximately 10,000 reads, which is comparable to the largest number of clusters, potentially resulting in sparse cluster abundances.

\begin{figure}[!htp]
   \begin{center}
   \begin{subfigure}[b]{0.49\textwidth}
        \centering
        \includegraphics[width=\textwidth]{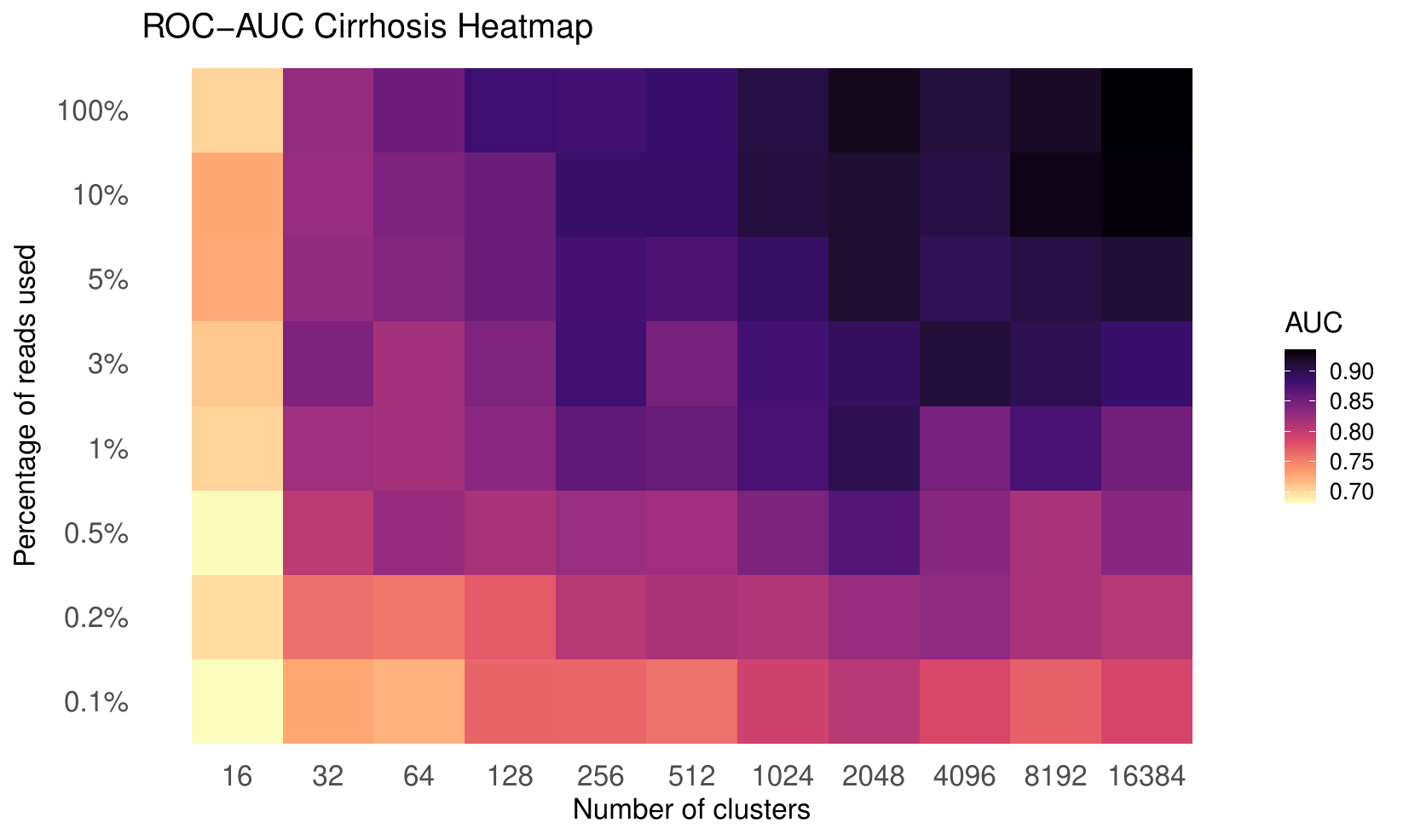}
        \caption{MetagenBERT-Glob perfomance on Cirrhosis regarding the proportion of reads used for assignment.}
    \end{subfigure}
    \hfill
    \begin{subfigure}[b]{0.49\textwidth}
        \centering
        \includegraphics[width=\textwidth]{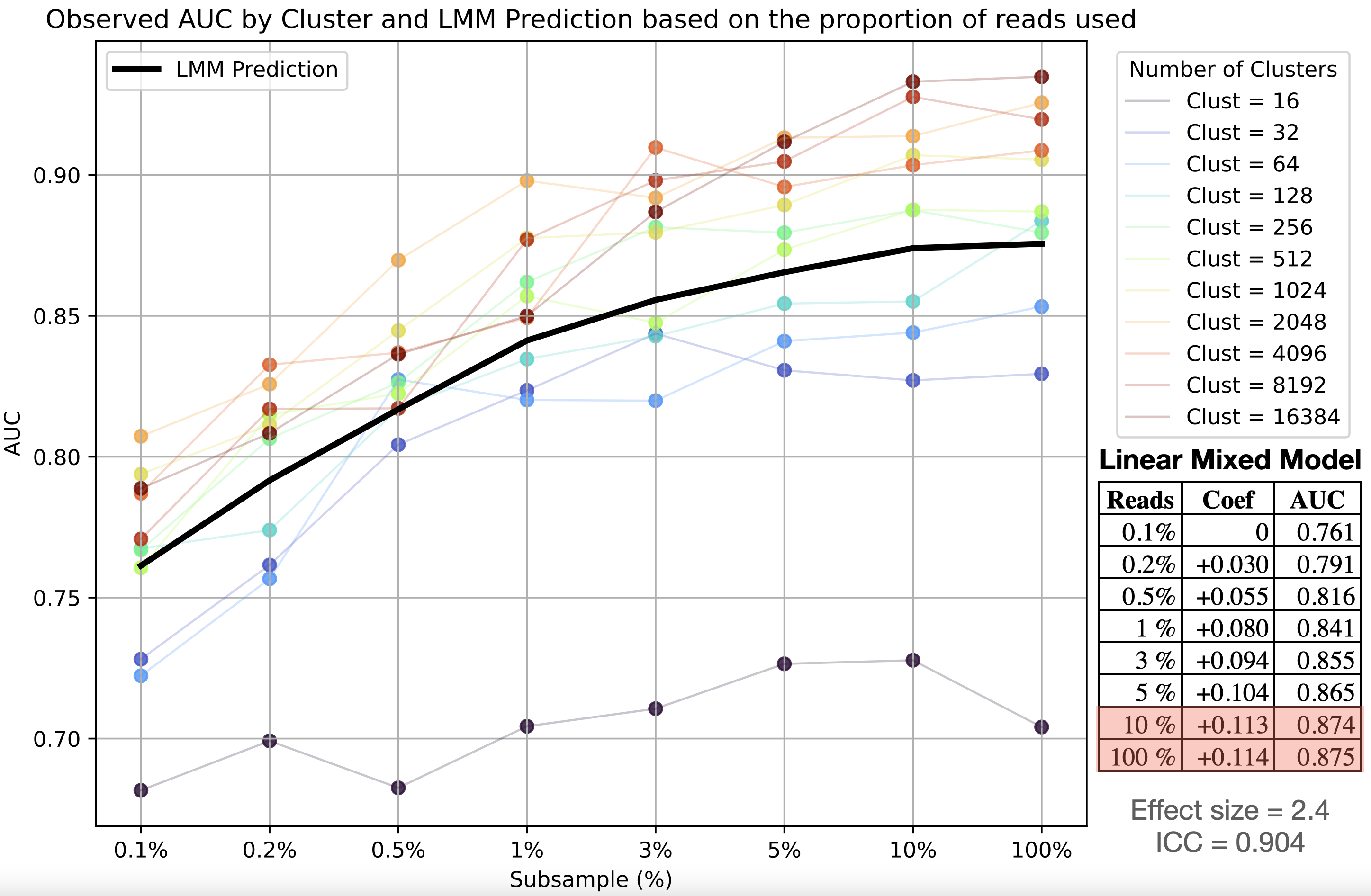}
        \caption{Linear Mixed Model for feature importance in Cirrhosis prediction}
    \end{subfigure}
    \hfill
   \end{center}
   \caption[\textbf{Comparison of Prediction performance when using MetagenBERT-Glob with different proportions of reads assigned.}]{\textbf{Comparison of Prediction performance when using MetagenBERT-Glob with different proportions of reads assigned.} The left panel of the figure shows performances heatmap according to different proportions of reads and number of clusters. Performances are comparable when using 10\% and 100\% of the data and start decreasing, especially for large number of clusters, under 3\% of reads and are unsatisfying under 0.5\% of reads. This results are confirmed on the right panel with the linear mixed model displaying mean performance with the evolution of reads proportion.}
   \label{fig:results_perc}
 \end{figure}

The results of this experiment are presented in Figure \ref{fig:results_perc}. Overall, similar trends are observed across all configurations. Using 10\% of the reads produces results nearly identical to those obtained with the full dataset, while 5\% yields slightly lower but still comparable performance. Below this threshold, performance begins to decline, particularly for configurations with a high number of clusters. When fewer than 1\% of reads are used, results are no longer comparable, even for smaller cluster numbers.

Notably, even sampling as few as one read out of every thousand still produces results substantially better than random, despite falling short of state-of-the-art performance.

Classification results indicate that using only a fraction of 10\% the data has minimal or no impact on performance.

In order to further assess this, we then trained a Linear Mixed Model (LMM) to test the effect of the variation of the proportion of reads on predictive performance. The results of this LMM can be seen in the right panel of Figure \ref{fig:results_perc}. We can see that, without surprise, the performance increase with the proportion of data used. However, the performance gain is becoming rather marginal as proportion of reads increases and is almost null when increasing from using 10\% of reads to 100\% of reads.

This result can be explained by the fact that, for these two configurations, the number of reads to assign is highly superior to the number of clusters. Therefore, the repartition of reads is similar in both configurations.

%% file: LaTeX/sections/3.Results/3.4.Clusters_Analysis.tex
Building on the results obtained for disease classification, we further investigated the structure of the embedding space and the biological relevance of the resulting clusters, with a particular focus on cirrhosis (analogous analyses for the other datasets are provided in the Supplementary Material).

\begin{figure}[htbp]
    \centering
        \begin{subfigure}[b]{0.33\textwidth}
         \centering
         \includegraphics[width=\textwidth]{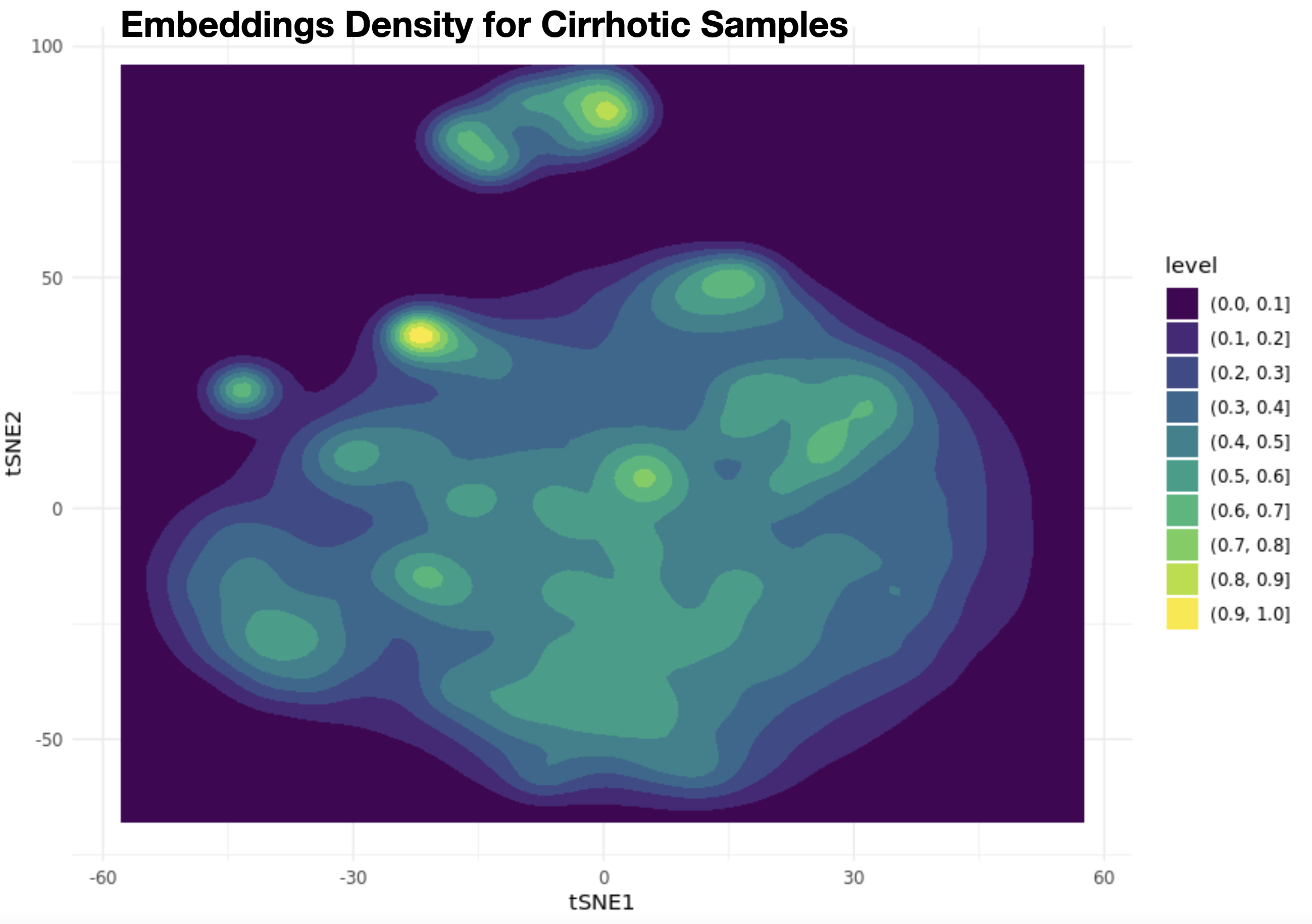}
     \end{subfigure}
     \begin{subfigure}[b]{0.33\textwidth}
         \centering
         \includegraphics[width=\textwidth]{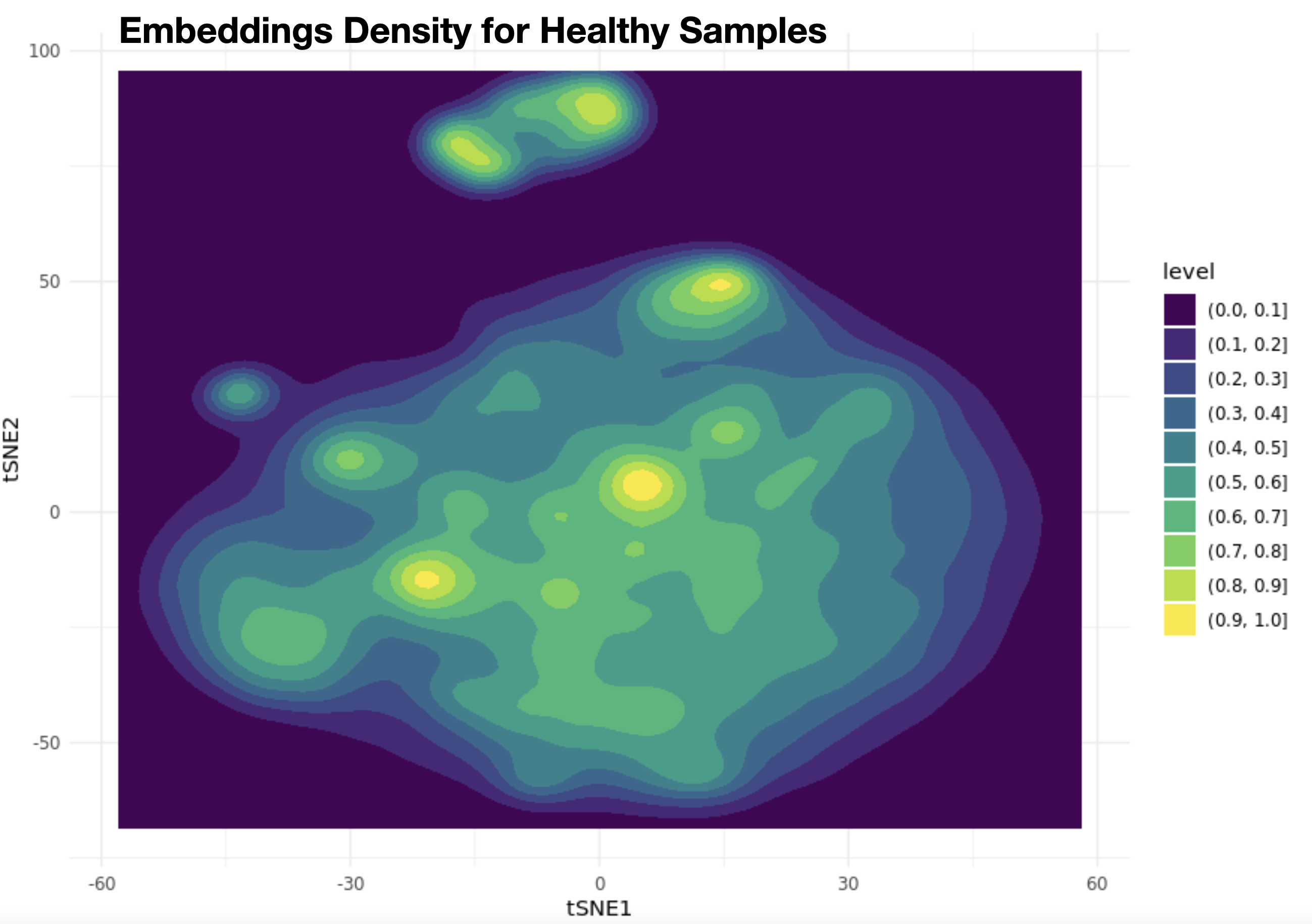}
     \end{subfigure}
     \begin{subfigure}[b]{0.33\textwidth}
         \centering
         \includegraphics[width=\textwidth]{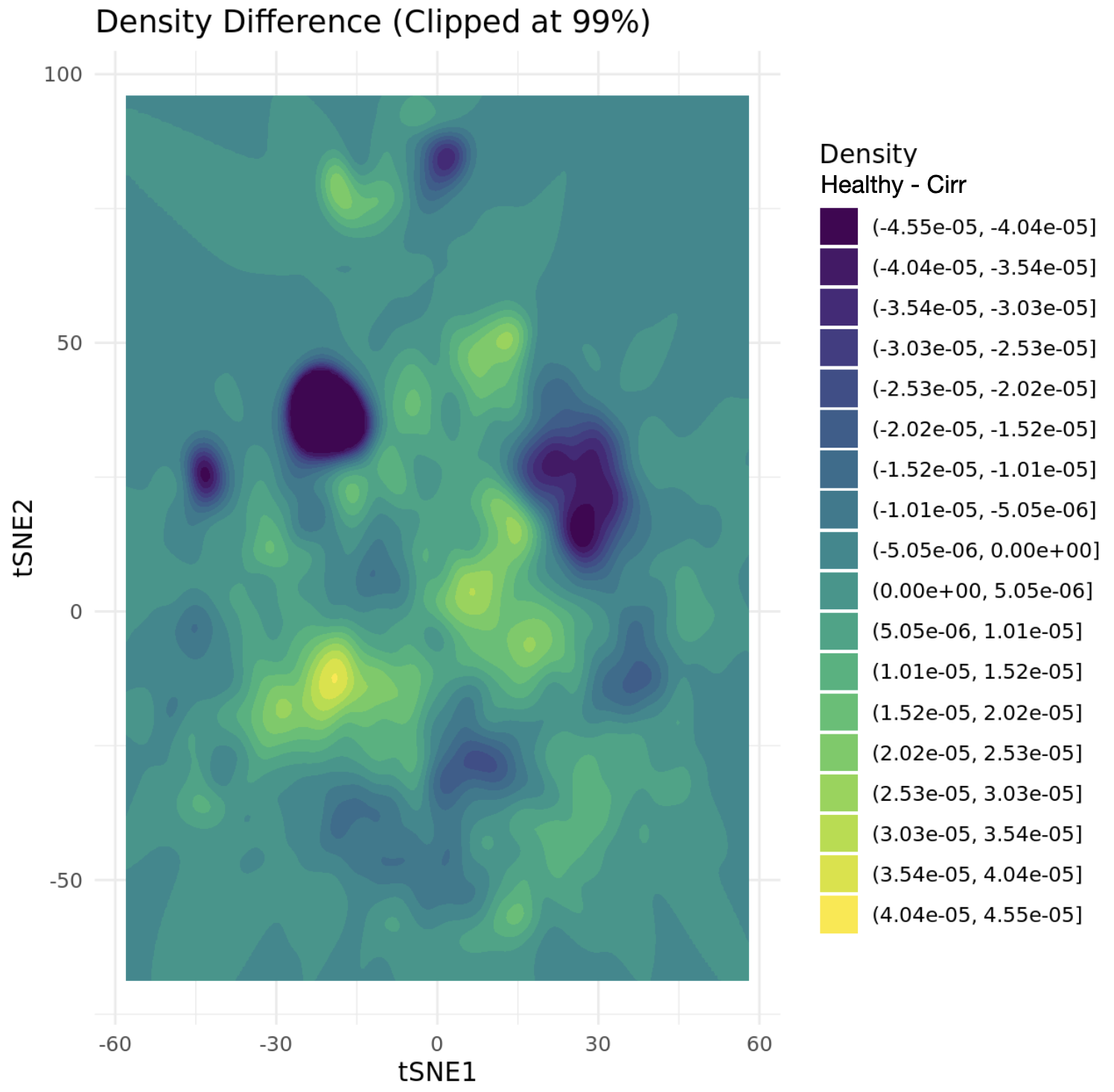}
     \end{subfigure}
    \caption[\textbf{T-sne projections of read embeddings from cirrhotic and healthy samples and density differences between both cases.}]{\textbf{T-sne projections of read embeddings from cirrhotic and healthy samples and differences of density between both cases.} The cirrhotic and case desities show a similar common pattern. However, cirrhotic samples display less density and most importantly, a highly represented cluster totally absent from case samples.}
    \label{fig:density}
\end{figure}

We first compared the latent representations of read embeddings between case and control samples. A subset of 400,000 embeddings drawn from both groups was used to compute a two-dimensional t-SNE projection, following an initial dimensionality reduction by principal component analysis (PCA) from 768 to 50 dimensions. Kernel density estimates of the resulting two-dimensional projections are reported in Figure \ref{fig:density}. As illustrated in the two leftmost panels, embeddings from cirrhotic and healthy samples share a common global structure, but also exhibit marked differences, as highlighted in the third panel. In particular, cirrhotic samples display a more heterogeneous density distribution and a highly concentrated cluster that is completely absent in healthy controls.

These observations indicate that disease-specific variations in embedding density and the emergence of distinct clusters can be captured by gLLM-based representations. They further suggest that clustering approaches that explicitly account for local density variations are particularly well suited for the analysis of metagenomic embedding spaces.

MetagenBERT embeddings have been shown to capture meaningful patterns relevant for microbiome classification. Furthermore, examining the evolution of classification performance when integrating these embeddings with species abundance suggests that the two representations encode distinct, potentially complementary, aspects of microbiome dynamics.

To further investigate this hypothesis, we explored the content of the embeddings and examined the potential biological interpretations of the resulting clusters.

\begin{figure}[htbp]
     
         \centering
         \includegraphics[width=\textwidth]{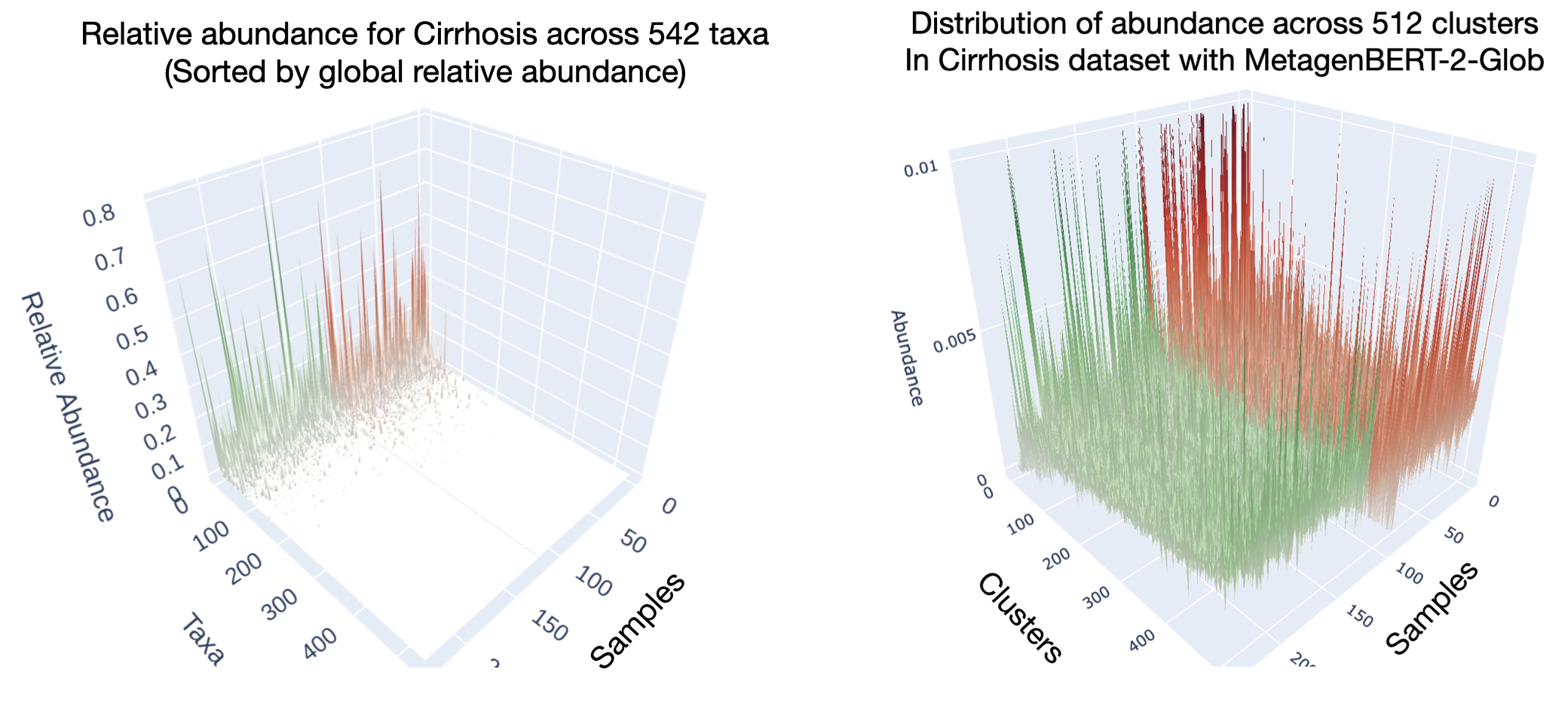}
     \hfill

     \caption[\textbf{Distribution of abundances of clusters from T2D,dataset using MetagenBERT-Glob-512 and species relative abundance in the same dataset.}]{\textbf{Distribution of abundances of clusters from T2D dataset using MetagenBERT-Glob-512 and species relative abundance in the same dataset.} The repartition profile is very different, with balanced clusters in the case of MetagenBERT representation and highly unbalanced sparse representation in species representations.}
     \label{fig:heatmap-all-foundation-512}
 \end{figure}

We first compared the abundance profiles derived from species-based and cluster-based representations, an example of which is shown in Figure \ref{fig:heatmap-all-foundation-512}. The two representations exhibited notable differences. Species-abundance profiles were dominated by a small number of highly abundant species, resulting in sparse and uneven distributions. In contrast, cluster-abundance profiles were considerably more balanced, with most clusters exhibiting comparable abundance levels. It should be noted that K-means clustering aims to minimize within-cluster variance rather than cluster size, identifying centroids that minimize the sum of squared distances between points and their assigned centers. Therefore, the relative uniformity observed in cluster abundances likely reflects variations in embedding density across samples rather than the influence of specific, highly abundant reads.

However, differences in the distribution of abundances do not necessarily imply that the clusters encode information distinct from that captured by species-level profiles. One could argue, for instance, that a single species might be partitioned across multiple clusters, yielding apparent differences in distribution without representing new information.

To evaluate this possibility, we assessed the correspondence between species-based and cluster-based profiles. Specifically, we computed pairwise distances between samples using the Aitchison distance \cite{aitchison}, a compositional-data adaptation of the Bray–Curtis distance \cite{bray-curtis}, thereby generating two distance matrices corresponding to species- and cluster-based profiles. We then applied a Mantel test to evaluate the correlation between these matrices. Results obtained using 512 clusters are presented in Table \ref{tab:mantel}, while additional results are reported in the Supplementary Material. The observed correlations were negligible, providing further evidence that MetagenBERT-derived profiles and taxonomic profiles capture distinct, potentially complementary aspects of microbiome structure.

\begin{table}[]
\resizebox{0.8\columnwidth}{!}{%
\begin{tabular}{l|l|l|l|l|l}
                        & Cirrhosis & Obesity & T2D    & CRC     & IBD    \\
                        \hline
Correlation Coefficient & 0.2346    & 0.0530  & 0.0076 & -0.0164 & 0.0897 \\
\hline
p-value                 & 0.0001    & 0.0001  & 0.7973 & 0.7912  & 0.2049
\end{tabular}%
}
\caption{\textbf{Table of correlation coefficients between species-based representations and MetagenBERT representations along with their p-values.}}
\label{tab:mantel}
\end{table}

\begin{figure}[htbp]
    \centering
    
        \includegraphics[width=\textwidth]{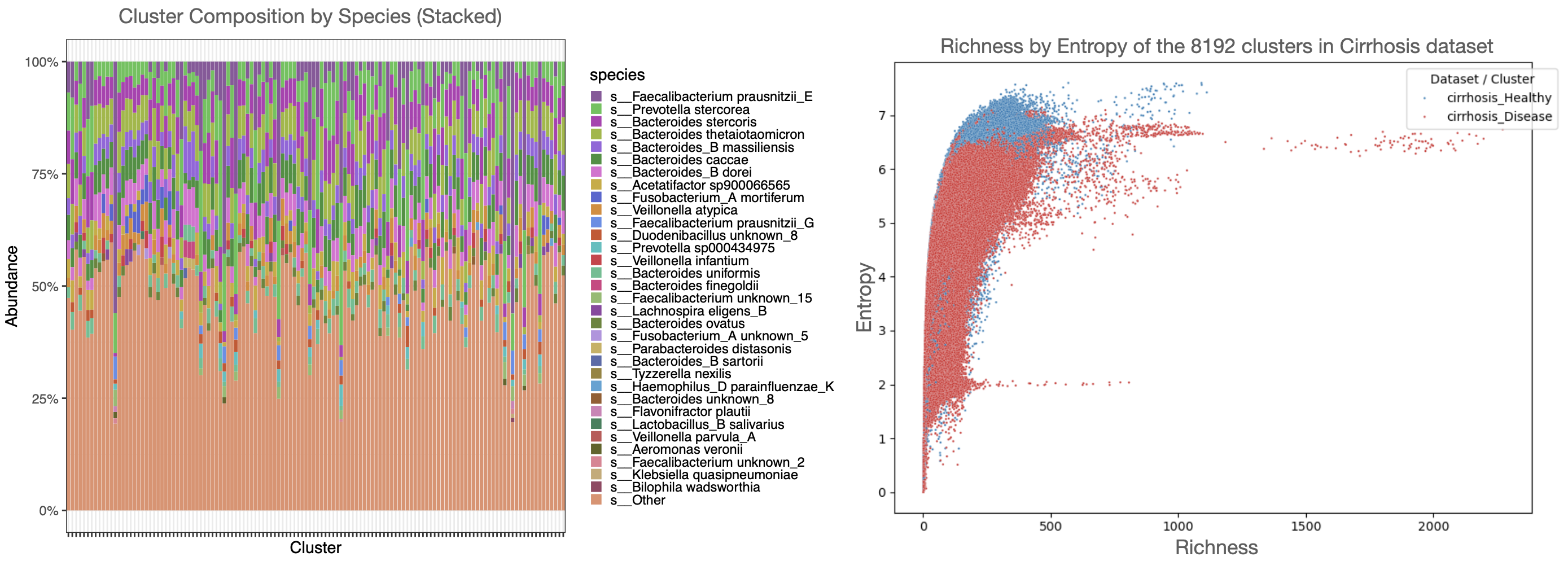}   
    \caption[\textbf{Clusters Species Compositions}]{\textbf{Clusters Species Compositions and entropy.} On the left panel, we see a barplot of the profile of species found inside each cluster of a sample, showing their highly diverse nature and the repartition of species in each clusters. On the right is visible the richness and entropy of each cluster in a dataset, further confirming that studied clusters contain numerous species in balanced proportions.}
    \label{fig:clust-comp}
\end{figure}

This claim is further supported by the composition of the clusters. For each cluster, reads assigned to it were analyzed using Centrifuge \cite{centrifuge} to retrieve their species of origin, based on the UHGG database. This allowed us to construct a taxonomic profile for each cluster. As shown in Figure \ref{fig:clust-comp}, the right panel presents a scatterplot of cluster richness and entropy across samples from the Cirrhosis dataset, indicating that clusters are highly diverse and contain many species in comparable proportions. The left panel illustrates, for a representative cirrhosis sample, the relative abundance of the most represented species within each cluster. These profiles are balanced across clusters, with no single species dominating, confirming the high intra-cluster diversity.

Together, these observations confirm that MetagenBERT-derived representations encode information distinct from conventional taxonomic profiles and could thus be used in conjunction with species-level analyses in metagenomic studies. The clusters identified may correspond to alternative biological groupings, such as functional guilds, or represent novel micro-communities not captured by standard taxonomic partitions, and even not real biological-based partitions, with clusters potentially covering different realities. Some clusters could reflect repeated genomic regions, important genes, or groups of species with shared functional roles. Further investigation of these clusters is warranted to elucidate the biological insights uncovered by LLM embeddings and to enhance both the interpretability and reliability of MetagenBERT representations.

\begin{figure}[htbp]
    \centering
    
        \includegraphics[width=\textwidth]{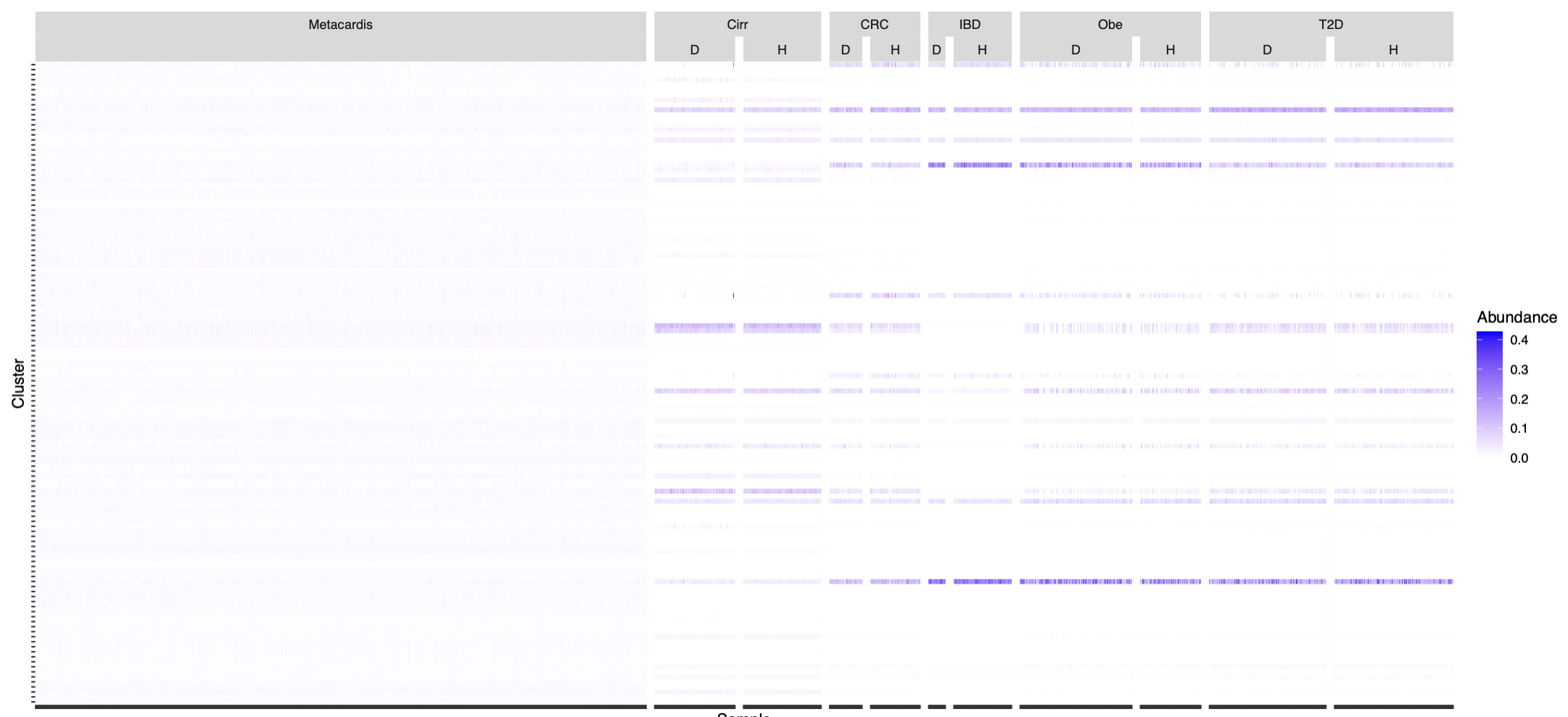}   
    \caption{\textbf{Heatmap of Metacardis cluster abundance and other benchmark datasets projections on Metacardis clustering.}]{\textbf{Heatmap of Metacardis cluster abundance and other benchmark datasets projections on Metacardis clustering.} We see here, for 128 clusters, the comparison between MetagenBERT-2-Glob clusters abundance obtained with Metacardis and MetagenBERT-2-Glob-Mcardis other benchmaks' projections on Metacardis.  We clearly see that Metacardis abundance is very balanced, as were abundances of each datasets when clustered using MetagenBERT-Glob, while their projections using MetagenBERT-Glob-Mcardis result in some specific highly populated clusters and a majority of them being almost to totally empty.}}
        \label{fig:dorsale}
\end{figure}

Next, we analyzed the cluster profiles obtained using our more general model, MetagenBERT-Glob-Mcardis (Figure \ref{fig:dorsale}). Compared to previous results, these profiles exhibited markedly different patterns. While Metacardis sample profiles remained relatively balanced, profiles from the other five cohorts were highly sparse, characterized by a few densely populated clusters alongside many near-empty clusters. This unexpected distribution may partly explain the lower predictive performance observed in Figure \ref{fig:foundation_results}. Notably, the most abundant clusters were shared across cohorts and did not show apparent differences between case and control samples, suggesting that the Metacardis dataset may be less suited to discriminate certain phenotypes. Potential explanations include technical factors, such as differences in sequencing platforms or preprocessing methods, as well as biological or population-level variation, which may lead benchmark datasets to populate peripheral or sparsely represented clusters within the Metacardis-defined space.

Indeed, each benchmark dataset tended to occupy a limited subset of clusters, which may produce less informative abundance profiles and introduce ambiguity across cohorts, ultimately constraining the generalization capacity of this transfer-based approach.

To evaluate whether classification models relied on consistent features across datasets, we examined the stability of predictive clusters. For each classification experiment, three independent training runs were performed, and the importance of the features for LASSO classification were recorded. Mean importance values across runs were computed to identify clusters that consistently contributed to predictions. Due to the sparsity-inducing nature of the LASSO penalty and the unbalanced cluster distributions, many features were assigned zero importance. Heatmaps displaying the relative importance of each feature across all cohorts and experimental configurations highlighted globally influential clusters while revealing distinct importance profiles for individual cohorts. This indicates that, although embeddings from different datasets largely occupy the same regions of the Metacardis-defined cluster space, the specific features discriminating case from control samples are disease-specific. In other words, the structural embedding framework supports diverse, cohort-specific discriminative patterns.

Overall, these findings demonstrate that MetagenBERT-Glob-Mcardis-derived representations can reliably distinguish healthy from diseased samples within individual cohorts, even in the absence of explicit model training on the specific phenotypes. This supports the feasibility of developing general-purpose metagenomic embedding models capable of producing transferable representations suitable for a wide range of clinical prediction tasks.

\begin{figure}[htbp]
    \centering
    \begin{subfigure}[b]{1\textwidth}
        \centering
        \includegraphics[width=\textwidth]{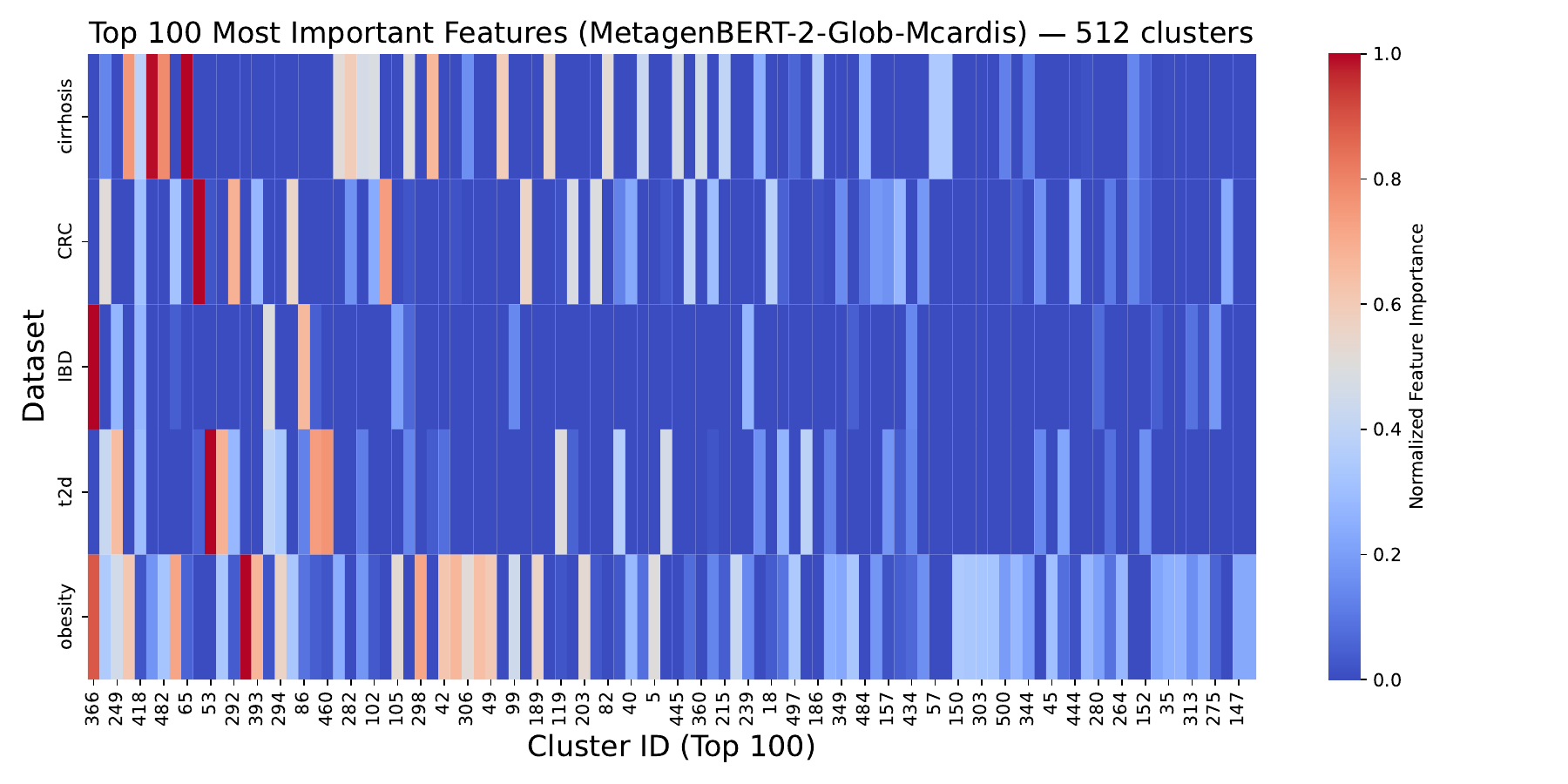}
        \caption{Top 100 features importance using MetagenBERT-2-Glob-Mcardis-512}
    \end{subfigure}
    \hfill
    \begin{subfigure}[b]{1\textwidth}
        \centering
        \includegraphics[width=\textwidth]{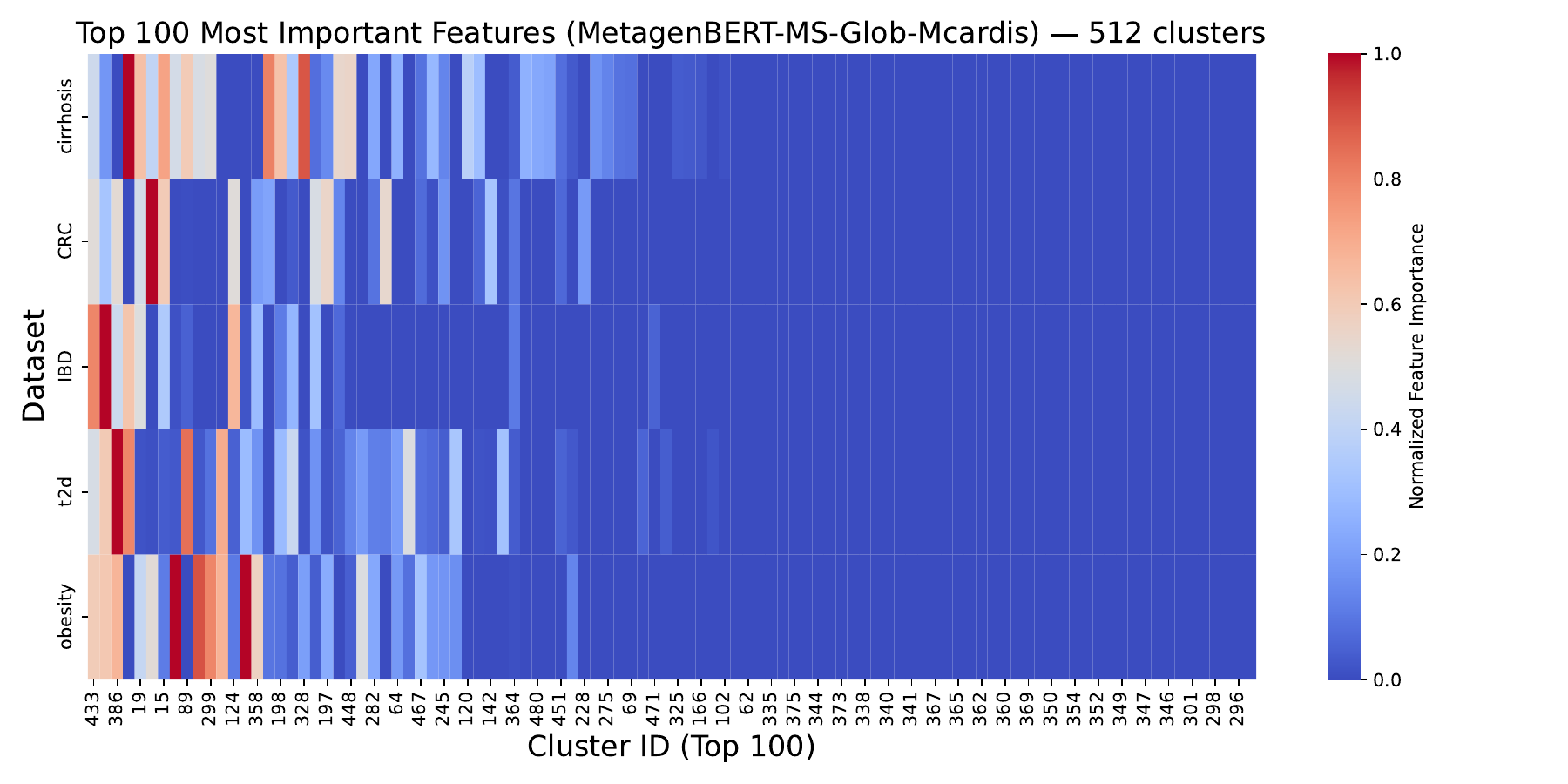}
        \caption{Top 100 features importance using MetagenBERT-MS-Glob-Mcardis-512}
    \end{subfigure}
        \hfill
    \caption[\textbf{Heatmap of importance of features from all datasets using MetagenBERT-Glob-Mcardis-512.}]{\textbf{Heatmap of importance of features from all datasets using MetagenBERT-Glob-Mcardis-512.} These features are the most important globally after cohort normalization Therefore, they are more likely to be shared among datasets. However, we see the five datasets follow very different profiles, showing the expressive power does not rely in the same clusters. This heatmap displays the importance in each dataset classification using the top 100 global feature importance. MetagenBERT-MS-Glob-Mcardis uses less features than MetagenBERT-2-Glob-Mcardis.}
    \label{fig:heatmap-assign-feature-512}
\end{figure}

%% file: LaTeX/sections/4.Discussion/4.Discussion.tex
Our approach, while yielding promising results, still presents several challenges, limitations, and opportunities for improvement.

\subsection{Limits in data}

Publicly available, well-annotated datasets are rare due to high costs, privacy constraints, and the complexity of collection, processing, and curation. A major recurring issue is the low number of samples relative to the extremely high dimensionality of the data, which limits model training, reduces cross-validation robustness, and leads to unstable performance estimates.

The Metacardis dataset partially alleviates this limitation by providing nearly 900 French samples and over 2,000 overall, but follow-up analyses still involve fewer than 200 patients per variable. Beyond sample size, dataset diversity is limited: microbiome composition is influenced not only by disease but also by lifestyle, diet, geography, medication (e.g., metformin) \cite{metformin}, and sequencing technology. 
We deem the latter to be a strong factor in the difficulties encountered to use MetagenBERT-Glob-Mcardis as a foundation model for other datasets. All these factors act as confounders and can introduce embedding biases, as different sequencing platforms may cluster together.

The authors therefore emphasize the importance of training on large, real, heterogeneous datasets. Ongoing initiatives such as the French Gut project and a global one-million-sample effort are expected to provide such data and offer strong potential for improving future foundation models like MetagenBERT-Glob.

\subsection{Choosing the right read embedder}

Although Transformer-based read embedding models such as DNABERT-2 perform strongly, they are trained on highly heterogeneous genomic data and may not optimally capture the specific characteristics of metagenomic reads. To address this, the authors developed DNABERT-MS, a continuously pre-trained version of DNABERT-2 on simulated metagenomic reads, with the goal of specializing the model for microbiome DNA.

DNABERT-MS showed improved performance at the read-embedding level, indicating that metagenomic specialization can enhance representation quality. However, these gains did not consistently translate into better disease-prediction performance, where DNABERT-MS and DNABERT-2 achieved similar results depending on the dataset. This suggests that while specialization is promising, larger, better-curated, and more biologically balanced metagenomic pretraining corpora are likely required to unlock its full benefit.

Moreover, some supplementary difficulties come with the use of short reads dataset, difficulties we tried to address with MetagenBERT-Glob.
Indeed, the shortness of the reads reduces the expressive power of each read, erasing long-term dependencies and losing potentially significant insight. On the other hand, their high number advocates for the use of quantitative methods as the clustering developed here. To this date, the structure of metagenomic datasets make their treatment through only one gLLM challenging.

The authors emphasize the rapid evolution of Transformer architectures, with new designs aiming to improve scalability, efficiency, and expressiveness—key needs for metagenomics, along with long read technologies \cite{long-reads} may bring up new opportunities in studying microbiome DNA through LLMs. 

Finally, the interpretability of Transformer embeddings remains a major open challenge, especially for short reads where learned features vastly outnumber observable tokens, making it difficult to link internal representations to concrete biological signals.

\subsection{Computational Cost}

The proposed approach remains computationally demanding, as embedding and clustering millions of reads requires substantial GPU resources, and the clustering step is also mildly time-consuming. Although experiments using only a fraction of reads show that the process can be simplified, large-scale deployment of language models for metagenomics is still costly.

The MetagenBERT-Glob framework partly mitigates this issue by shifting most of the computational burden to the pre-deployment phase. Once clustering model is trained, processing a new sample only requires embedding a subset of reads (e.g., 10\%), assigning them to precomputed clusters, and computing abundance vectors, which are relatively fast operations. Thus, the heavy computation is largely removed from the point-of-care setting.

Even so, the optimized pipeline remains non-trivial in terms of compute. The authors suggest future work on lighter embedding models and embedding quantization to further reduce inference time and make near-real-time metagenomic analysis more feasible.

\subsection{Clustering Method}

The authors highlight the extreme complexity of their embedding space, where each metagenome contains tens of millions of reads represented in a 768-dimensional space. Although K-Means was chosen for its simplicity and scalability, its reliance on Euclidean distance makes it ill-suited to high-dimensional settings due to the curse of dimensionality, which degrades distance discriminability and can harm clustering quality.

To overcome this, the authors propose exploring alternative high-dimensional clustering strategies. These include density-based methods such as HDBSCAN \cite{malzer_hybrid_2020}, which better handle heterogeneous local densities, as well as subspace clustering algorithms like CLIQUE and PROCLUS \cite{noauthor_automatic_nodate} \cite{aggarwal_fast_nodate} that aim to discover structure in meaningful low-dimensional subspaces. Spectral clustering \cite{ng_spectral_nodate} combined with dimensionality reduction (PCA or UMAP) is also suggested, although preliminary PCA experiments showed strong information loss when reducing dimensionality from 768 to 256 or 64 dimensions.

\subsection{Cluster interpretability}

Although MetagenBERT-Glob clusters provide an effective alternative to species-level abundances for disease classification, their biological interpretation remains difficult. The clusters encode information distinct from taxonomic profiles, but their high dimensionality, low variance after reduction, and similar abundance distributions hinder visualization and intuitive understanding of the embedding space.

Nevertheless, clustering proved robust: repeating the procedure on the same data or with different read subsampling strategies produced highly consistent clusters, indicating that they capture meaningful, stable patterns rather than noise.

The authors argue that future work should focus on interpretability, for instance by linking clusters to other microbial pathways, and by examining the most influential features selected by the LASSO classifier. Improving transparency is essential for clinical adoption, as it would connect abstract embeddings to concrete biological mechanisms and actionable insights.

\subsection{Conclusion}

With MetagenBERT-Glob, we successfully developed a new method for gut microbiome samples embedding directly from DNA sequences without external information or reliance on pre-computed catalogs. We leveraged gLLM expressive power to represent this DNA in a 768 dimensions space and created a method to partition this space and create a reference separation for all samples studied. We showed that this embedding allowed for efficient classification through simple machine learning models on five benchmark datasets, competing with State of the Art models.
Moreover, we showed that this representation could be used in conjunction with taxonomic profiles and began to explore the contents of the clusters obtained with the computed partition.
We believe our method can serve for metagenomic exploration and representation and that, with extensive training on diverse, carefully curated datasets, MetagenBERT-Glob can generalize and produce reliable microbiome representations.